\documentclass[preprintnumbers,prd,onecolumn,floatfix,superscriptaddress, nofootinbib]{revtex4-2}

\usepackage{setspace}
\usepackage{graphicx}
\usepackage{subfig}
\usepackage{epsfig}
\usepackage{bm}
\usepackage{amssymb}
\usepackage{float}
\usepackage{amsmath}
\usepackage{dcolumn}
\usepackage[colorlinks]{hyperref}
\usepackage[usenames,dvipsnames]{color}
\usepackage{enumitem}

\hypersetup{ breaklinks=true, pdfstartview={FitH}, colorlinks=true, linkcolor=blue, citecolor=red, filecolor=magenta, urlcolor=blue, anchorcolor=green, linktocpage=true }

\def\doi{http://doi.org}

\newcommand{\disp}{\displaystyle}
\newcommand{\be}{\begin{equation}}
\newcommand{\ee}{\end{equation}}
\newcommand{\beano}{\begin{eqnarray*}}
\newcommand{\eeano}{\end{eqnarray*}}
\newcommand{\ba}{\begin{eqnarray}}
\newcommand{\ea}{\end{eqnarray}}
\newcommand{\no}{\nonumber}

\newcommand{\htwo}{\mbox{\hspace{2em}}}

\begin{document}

\title{Dynamics of a parametrized dark energy model in $ f(R,T) $ gravity}

\author{J. K. Singh}
\email{jksingh@nsut.ac.in}
\affiliation{ Department of Mathematics,
 Netaji Subhas University of Technology\\
 New Delhi-110078, India}
 
\author{Akanksha Singh}
\email{akanksha.ma19@nsut.ac.in}
\affiliation{ Department of Mathematics,
 Netaji Subhas University of Technology\\
 New Delhi-110078, India}
 
\author{G. K. Goswami}
\email{gk.goswami9@gmail.com}
\affiliation{ Department of Mathematics,
 Netaji Subhas University of Technology\\
 New Delhi-110078, India}
 
\author{J. Jena}
\email{jjena67@rediffmail.com}
\affiliation{ Department of Mathematics,
 Netaji Subhas University of Technology\\
 New Delhi-110078, India}

\begin{abstract}
	
\begin{singlespace}

\noindent
We investigate a flat FLRW-model in $ f(R, T) $-gravity, which includes the quadratic variation in scalar curvature $ R $ and the linear term of the trace of the stress-energy tensor $ T $. In turn, we establish the model has the behaviour of the late time Universe, which is accelerated expanding. Using the parametrization of scale factor $ a(t) $, we propose a model, which begins with point-type singularity, \textit{i.e.}, the model starts with a point of zero volume, infinite energy density and infinite temperature. The model's behaviour is accelerated expanding at present and $ \Lambda$CDM in late times. Finally, the proposed model behaves like a quintessence dark energy  model in the present time and is consistent with standard cosmology $ \Lambda $CDM in late times.
\end{singlespace}

\end{abstract}

\maketitle

\noindent
PACS numbers: {04.50.-h, 98.80.-k.}\\
Keywords: {FLRW Universe, $ f(R,T) $-gravity, parametrization, dark energy, statefinder diagnostic}

\section{ Introduction}

\qquad Observational data indicating high redshift supernovae emphasises the rising belief of late-time cosmic acceleration. This phenomenon also gets its support from the observations related to weak lensing, microwave background and large scale structure. A current problem of modern cosmogenesis is knowing what produces the repulsion during cosmic expansion. The phenomena can theoretically be explained by either adding an unusual matter component with substantial negative pressure to the energy-momentum tensor \cite{Sahni:1999gb,Sahni:2006pa,Copeland:2006wr,Linder:2008pp,Caldwell:2009ix,Silvestri:2009hh,Frieman:2008sn} or changing gravity itself. 

The most straightforward dark energy candidate is the cosmological constant. The cosmological constant is equivalent to a fluid having an equation of state $ \omega = -1$,  where $ \omega $ is the ratio of pressure that dark energy puts on the Universe to the energy per unit volume. Traditionally, Einstein added the cosmological constant $ \Lambda $ into the field equation for gravity for the reason that it allows for a closed, static, finite Universe in which the geometry is determined by the matter's energy density \cite{Padmanabhan:2002ji}. Einstein thought that ordinary matter would be required to bend geometry, a necessity that, according to him, was strongly tied to the Mach's principle. This optimism was quickly dashed when de Sitter discovered a solution to Einstein's equations with a cosmological constant including no matter \cite{deSitter:1917zz}. Despite two seminal studies by Friedmann and one by Lemaitre \cite{Friedman:1922kd,Friedmann:1924bb,Lemaitre:1927zz}, most workers were resistant to the concept of an expanding cosmos. Lemaitre's work helped the community accept the idea of expanding the entire Universe. The cosmological constant has a tumultuous history and has frequently been embraced or dismissed for erroneous or inadequate reasons. The steady-state model was the first cosmological model that used the cosmological constant as a fundamental quantity  \cite{Bondi:1948qk,Hoyle:1948zz,Hoyle:1964}. It took advantage of the fact that a Universe with a cosmological constant has the property of time translational invariance in a given coordinate system. Unlike current cosmology, which readily conjures negative energies or pressure, largely abandoned steady-state cosmology was with the discovery of CMBR.

A remarkable result in modern cosmology, that is, the presence of cosmological constant fueling the Universe's current acceleration, has been growing steadily, as demonstrated in \cite{SupernovaCosmologyProject:1998vns,SupernovaSearchTeam:1998fmf,Riess:1998dv}. It was recently discovered that the enigmatic dark energy governs the late-time dynamics of the current accelerating cosmos. According to the interpretation of the astrophysical findings, such dark energy fluid (assuming it is fluid) is distinguished by negative pressure, and its equation of state parameter $ \omega $ is approaching $ -1 $ \cite{Nojiri:2005vv}. Mamon \textit{et al.} has discussed a unified model of dark matter and dark energy \cite{AlMamon:2021sgv}. Numerous studies cover various areas related to $ f(R) $ gravity and accompanying cosmic dynamics. Alternative gravitational theories represented by Lagrangians based on distinct general functions of the Ricci scalar have been shown to create lucid theoretical models to describe the experimental evidence of the Universe's acceleration. Allemandi \textit{et al.} continued this examination of cosmological possibilities of alternative gravity theories that rely on (other) curvature invariants \cite{Allemandi:2004wn}. A comprehensive investigation of the dynamics of $ R^n $ gravity cosmological models is discussed in \cite{Carloni:2004kp}. Viable options other than dark energy for explaining late-time cosmic acceleration can be found in modified gravitational theories. Also, few instances of discussions on low-curvature Einstein-Hilbert action corrections are noticed. Such frameworks feature unstable de Sitter solutions in general and, based on the theory's characteristics, can have late-time accelerating attractor solutions \cite{Easson:2004fq}. Higher-order gravity theories have recently gained much attention as potential options for explaining observable cosmic acceleration without the requirement for a scalar field. 

The function $ f(R) $ of the Ricci scalar curvature entering the gravitational Lagrangian and regulating the evolution of the Universe is a crucial component \cite{Capozziello:2005ku}. The fundamental goal of the work in \cite{Rahaman:2020dgv} was to examine the existence of compact spherical systems that represent anisotropic matter distributions in the context of alternative gravity theories, notably the $ f(R,T) $ gravity theory. The goal of this study in \cite{Shabani:2017rye} is to revamp a particular class of $ f(R,T) $ gravity models in which the Einstein-Hilbert action is supported by an arbitrary function of the energy-momentum-tensor's trace without losing its generality. Rosa introduced an alternative scalar-tensor representation in $f(R,T)$ gravity \cite{Rosa:2021teg}. Gonclaves et al. used reconstruction methods to obtain cosmological solutions in the scalar-tensor representation of $f(R,T)$ gravity \cite{Goncalves:2021vci}. Quite a number of researchers have already executed impressive work in $ f(R,T) $ gravity  \cite{Singh:2014kca,Singh:2015hva,Jamil:2011ptc,Singh:2015bzm,Maurya:2019iup,Sahoo:2017poz,Nagpal:2018mpv,Singh:2017qls,Singh:2014oga,Pawar:2021sro,Maurya:2019sfm,Harko:2011kv,Starobinsky:1980te}. Mishra et al. have reconstructed cosmological models using hybrid scale factor \cite{Mishra:2017zce, Mishra}.

General relativity's generalization can be used as a feasible explanation of the early-time inflation with late-time cosmic acceleration. The various representations of a multitude of modified theories, such as traditional $ F(R) $ and Ho$ \check{r} $ava-Lifshitz $ F(R) $ gravity, scalar-tensor theory, Gauss-Bonnet theory etc., as well as relationships between them, are studied in \cite{Nojiri:2010wj}. Nojiri \& Odintsov \cite{Nojiri:2004bi} propose that the expansion of the cosmos may result in the dominance of dark energy over standard matter. The effective quintessence aptly describes the existing cosmic speed-up. $ F(R) $, $ F(G) $ and $ F(T) $, are a few standard modified gravity theories. The objective of \cite{Nojiri:2017ncd} is to highlight all the relevant details on inflation, dark energy and bouncing cosmologies through multiple modified gravity models. The open irreversible thermodynamic interpretation of a simple cosmological model is described in detail for the $ f(R,T) $ gravity theory \cite{Harko:2014pqa}. In \cite{Harko:2010zi}, the cosmological implications of the non-minimal coupling matter-geometry coupling were thoroughly explored. In modified $ f(R) $ gravity models, the type of the coupling entirely and distinctively determines both the matter Lagrangian and the energy-momentum tensor.
	
The paper is organised as follows.  Section II provides a succinct treatment of $ f(R,T) $ theory. We get highly non-linear field equations by taking the $ f(R,T) $ as a blend of a $ R $-dependent part having terms up to the quadratic power of $ R $ and a linear $ T $-dependent part. We apply an ansatz for the scale factor $ a(t) $ to find the solution of the field equations and to investigate the behaviour of the geometrical parameters such as the Hubble parameter $ H(t)$ and the deceleration parameter $ q(t) $. The numerical results obtained for the energy density $ \rho $, fluid pressure $ p $ and the ratio of pressure that dark energy puts on the universe to the energy per unit volume $ w $ for the proposed model are depicted for their interpretations. The energy conditions are examined to analyze and interpret the resulting solution. In section III, we exhibit the jerk parameter, snap parameter, the lerk parameter, Om diagnostic, the velocity of sound and statefinder diagnostic tools graphically to test the correctness of our model. Section IV covers various cosmological tests for determining distances in cosmology using the parametrization above. Section V contains concluding remarks on the works presented in the manuscript.

\section{ FLRW space-time in $ f(R,T) $-gravity}

\qquad The Einstein field equations are given by 
\ba \label{1}
&& R_{ij}-\frac{1}{2} R g_{ij} = -\frac{8\pi G}{c^4}T_{ij},
\ea
where $ R_{ij} $ is the Ricci tensor, $ R $ the Ricci scalar, $ g_{ij} $  the covariant metric tensor of order 2, $ G $ the gravitational constant, $ c $ the speed of the light, and $ T_{ij} $  the energy-momentum-tensor. The energy-momentum-tensor on the right-hand side of Einstein's equations represents the contribution of the Universe's matter component, whereas the left-hand side symbolises pure geometry. There are two ways to cause an accelerated expansion: either $ (i) $ by adding an unusual type of matter to the energy-momentum-tensor, such as a cosmological constant or a scalar field, or $ (ii) $ by evolving the geometry itself. Geometrical alterations can be caused by quantum phenomena, such as the increased curvature modifications in the Einstein–Hilbert action \cite{Copeland:2006wr}. Gravitational field equations in $ f(R,T) $ gravity are derived from the following action
	
\ba\label{2}
&& S= \frac{1}{16\pi G} \int (f_1(R)+2f_2(T))\sqrt{-g} d^4x + \int S_m \sqrt{-g} d^4x ,
\ea
where $ f_1(R) $ and $ f_2(T) $ are arbitrary functions of their respective arguments and $ S_m $ is the matter Lagrangian density.  The energy-momentum-tensor will be derived from variation of second term of (\ref{2}). 

 Varying the action in equation (\ref{2}) with respect to  $ g_{ij} $ leads to the following equation
\ba \label{3}
&& f_1'(R) R_{ij}-\frac{1}{2} g_{ij} (f_1(R)+2f_2(T))+(g_{ij}\Box-\nabla_i \nabla_j)f_1'(R)=8\pi G\, T_{ij}-2f_2'(T)(T_{ij}+\Theta_{ij}),
\ea
where $( ' )$ represents the ordinary differentiation with respect to the variable in the argument, and other symbols have their usual meanings \cite{Nagpal:2019vre}. 

The energy-momentum-tensor for a perfect fluid is given by 
\ba \label{4}
&& T_{ij}=(\rho +p)u_{i}u_{j}-p g_{ij},
\ea
wherein $ \rho $ symbolises the energy density and $ p $ is the fluid's pressure which is there in the Universe. Here, the four-velocity vector $ u_i=(0,0,0,1) $ satisfies $ u^iu_i=1 $ and $ u^i \nabla_j u_i =0 $ in a comoving coordinate system. 

The gravitational equation of motion (\ref{3}) can also be written as
\ba\label{5}
&& f_1'(R) R_{ij}-\frac{1}{2}g_{ij} (f_1(R)+2f_2(T))+(g_{ij}\Box-\nabla_i \nabla_j) f_1'(R)= 8\pi G T_{ij}+2f_2'(T)(T_{ij}+p g_{ij}).
\ea
Equation (\ref{5}) can be represented in a more standard form as follows
\ba\label{6}
&& G_{ij}=R_{ij}-\frac{1}{2}R g_{ij} =\frac{8\pi G}{f_1'(R)}(T_{ij}+T_{ij}^{'}),
\ea
where
\ba
&& \disp T_{ij}^{'}=\frac{1}{8\pi G}\Big(\frac{1}{2} g_{ij}((f_1(R)+2f_2(T))-R f_1'(R))+(\nabla_{i} \nabla_{j} -g_{ij} \Box) f_1'(R)+(T_{ij}+pg_{ij})2f_2'(T)\Big). \no
\ea 
 It may be noted that by using $ \alpha=0 $ and $ \lambda=0 $ in the field equations (\ref{6}), one can obtain the Einstein's Field Equations for General Relativity.\\\\
 We considered $ f(R,T) $ gravity following FLRW space-time which represents a spatially homogeneous and isotropic Universe.  Particular forms of $ f_1(R) $ and $ f_2(T) $  are considered for further study, which are as per the following
\ba \label{7}
 && f_1(R)=R+\alpha R^2, \htwo  f_2(T)=\lambda T. 
\ea
The model in the present work has a dynamical variable in the form of the scale factor $ a(t) $ \cite{Bolotin:2015dja}. The following spatially flat FLRW line element will be used to study this model
\ba\label{8}
&& ds^{2}=dt^{2}-a^{2}(t)(dx_1^{2}+dx_2^{2}+dx_3^{2}).
\ea
 Considering $ G=1 $, the trace $ T $ of the energy-momentum-tensor $ T_{ij} $ and Ricci scalar curvature $ R $ are given by
\ba\label{9}
&& T=\rho -3p, \htwo R=-6(2H^2+\dot{H}),
\ea
where $ H=\dot{a}(t)/{a(t)} $ represents the Hubble parameter and  the dot over $ a $ signifies the differentiation with respect to cosmic time $ t $. In view of the equations (\ref{4}), (\ref{6}), (\ref{7}), (\ref{8}) and (\ref{9}) we obtained the following forms of field equations
\ba \label{10}
&& \left(8\pi+3\lambda \right)\rho-\lambda p=3H^{2}+18\alpha\left(\dot{H}^{2}-6H^{2}\dot{H}-2H\ddot{H}\right), \no\\
&& \left(8\pi+3\lambda \right) p-\lambda \rho=-2\dot{H}-3H^2+6\alpha\left(26\dot{H}H^2+2\dddot{H}+14H\ddot{H}+9\dot{H}^2\right).
\ea 
Solving equations  in (\ref{10}) we get  the following value for $\rho $ and $ p $
\ba\label{11}
&& \rho=\frac{\lambda\left(-2\dot{H}-3H^2+6\alpha\left(26\dot{H}H^2+2\dddot{H}+14H\ddot{H}+9\dot{H}^2\right)\right)+\left(8\pi+3 \lambda \right)\left(3H^{2}+18\alpha\left(\dot{H}^{2}-6H^{2}\dot{H}-2H\ddot{H}\right) \right)}{\left(8\pi+2 \lambda \right) \left(8\pi+4 \lambda \right)}, \no\\
&& p=\frac{\left(8\pi+3 \lambda \right)\rho-3H^{2}-18\alpha\left(\dot{H}^{2}-6H^{2}\dot{H}-2H\ddot{H}\right)}{\lambda }.
\ea

The effective energy density $ \rho_{eff} $ and effective pressure $ p_{eff} $ are computed by equations in (\ref{10}) as:
\ba\label{12}
&&
\rho_{eff} = \frac{3\left(n+\beta t\right)^2}{8\pi t^2}, \no\\
&&
p_{eff} = \frac{\left(2-3n\right)nt^2 - 12n\left(12 + n\left(26n-37\right)\right)\alpha - 6nt\left(t^2 + 8\left(13n-7\right)\alpha\right)\beta - 3t^2\left(t^2+104n\alpha\right)\beta^2}{8\pi t^4}.
\ea

It may be noted that once the parametrized form of either $ a(t) $ or $ H(t) $ is determined, the values of $ \rho $ and $ p $ can be obtained from (\ref{11}). There are number of parametrization patterns are available in the literature which give analytical solution to the field equations and can be termed as independent method to explore various dark energy models. We consider the following parametrized form of the cosmic scale factor $ a $. The motivation behind such hybrid form is that first the power law term probably gives the deceleration era of the universe, while the second exponential term produces the late time acceleration. The exponential term gives a high rate of expansion, the power-law term restricts it causing moderate accelerating expansion at present. The hybrid scale factor induces transitional behavior from deceleration to acceleration \cite{Nojiri:2022xdo, Odintsov:2021yva, Akarsu:2013xha}.

\ba \label{13}
a(t)= t^n \exp(\beta t),
\ea
wherein we have arbitrary constants $ \beta $ and $ n $, $ n>0 $. From equation (\ref{13}), the Hubble parameter $ H(t) $ and the deceleration parameter $ q(t) $ can be determined as
\ba \label{14}
&& H(t)= \beta + \frac{n}{t}, \htwo q(t)= -1 + \frac{n}{(\beta t + n)^2}.
\ea
From equation $ (\ref{14})_2 $, the correlation between model parameters $ \beta $ and $ n $ can be established as
\ba \label{15}
&& \beta t_0 = \Big(\frac{n}{q_0 +1}\Big)^\frac{1}{2} -n,
\ea
where $ t_0 $ signifies the current time and $ q_0 $ signifies the current value of deceleration parameter. In relation to discrete values for model parameter $ n $, one can obtain discrete values for other model parameter $ \beta $ when $ t_0 $ and $ q_0 $ are taken into account. In this present work, we  considered the values $ t_0=13.8 $ and $ q_0=-0.54 $ \cite{Mamon:2016dlv}

\begin{figure}\centering
	\subfloat[]{\label{a}\includegraphics[scale=0.45]{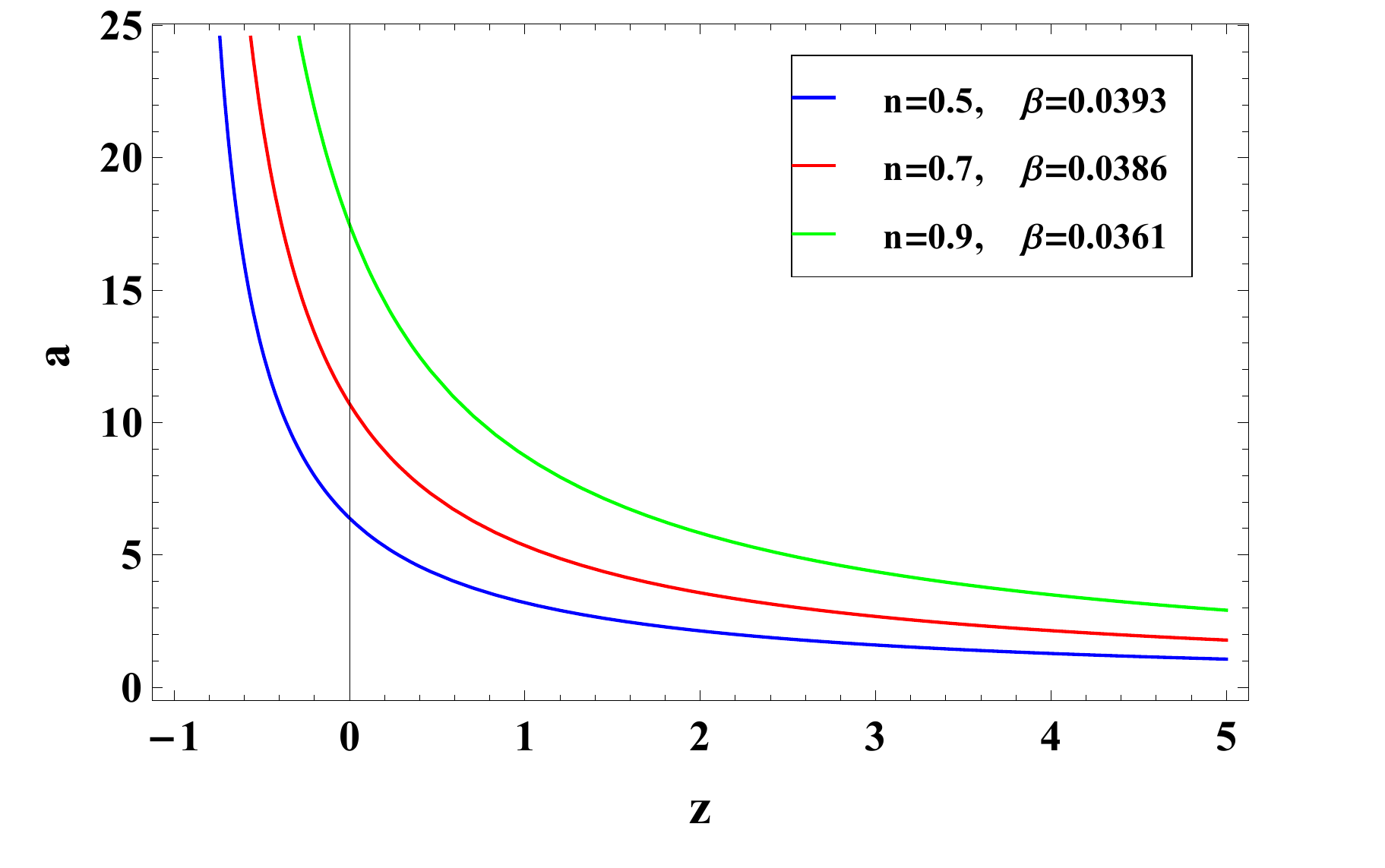}}\hfill
	\subfloat[]{\label{b}\includegraphics[scale=0.45]{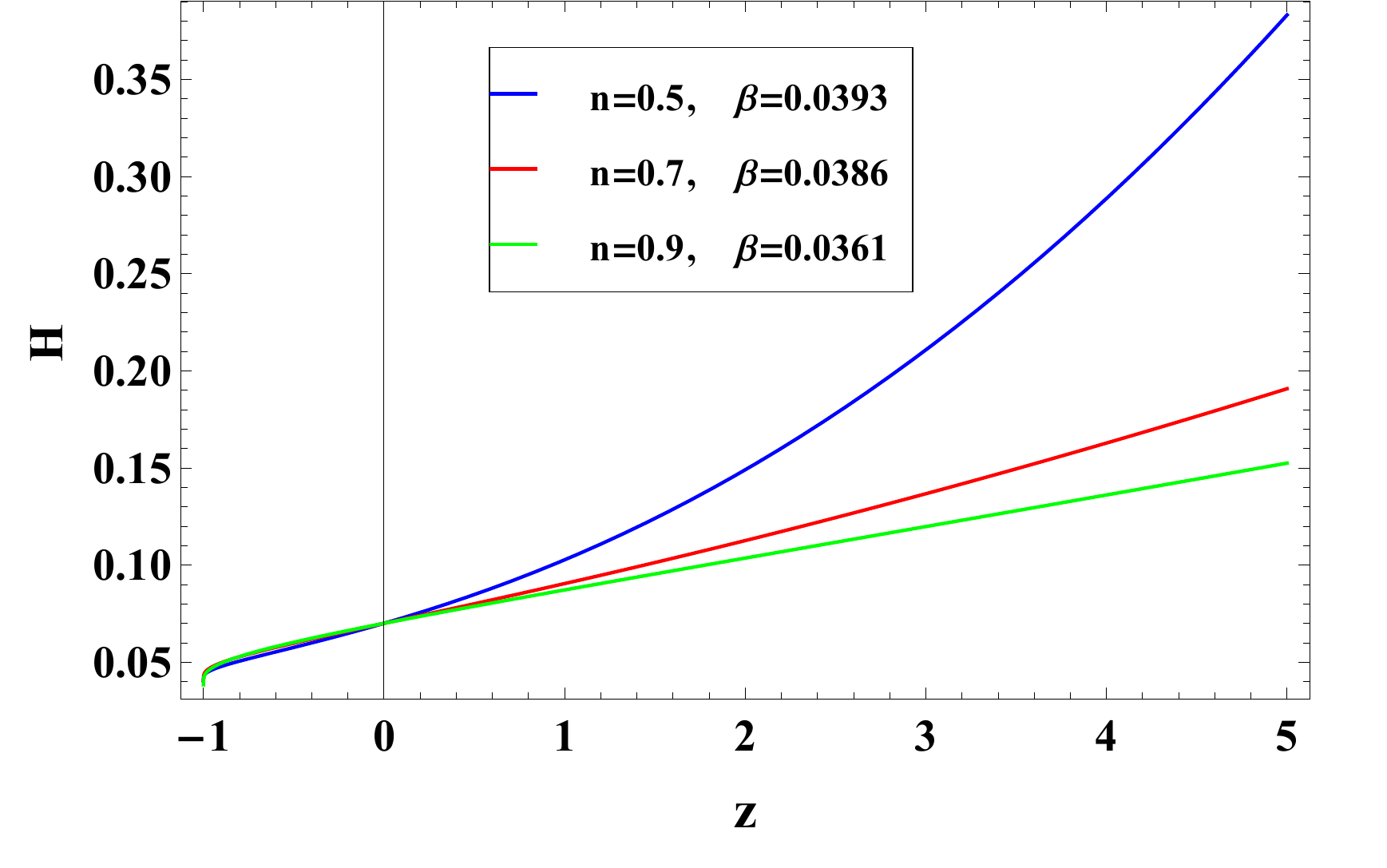}}\par 
	\subfloat[]{\label{c}\includegraphics[scale=0.45]{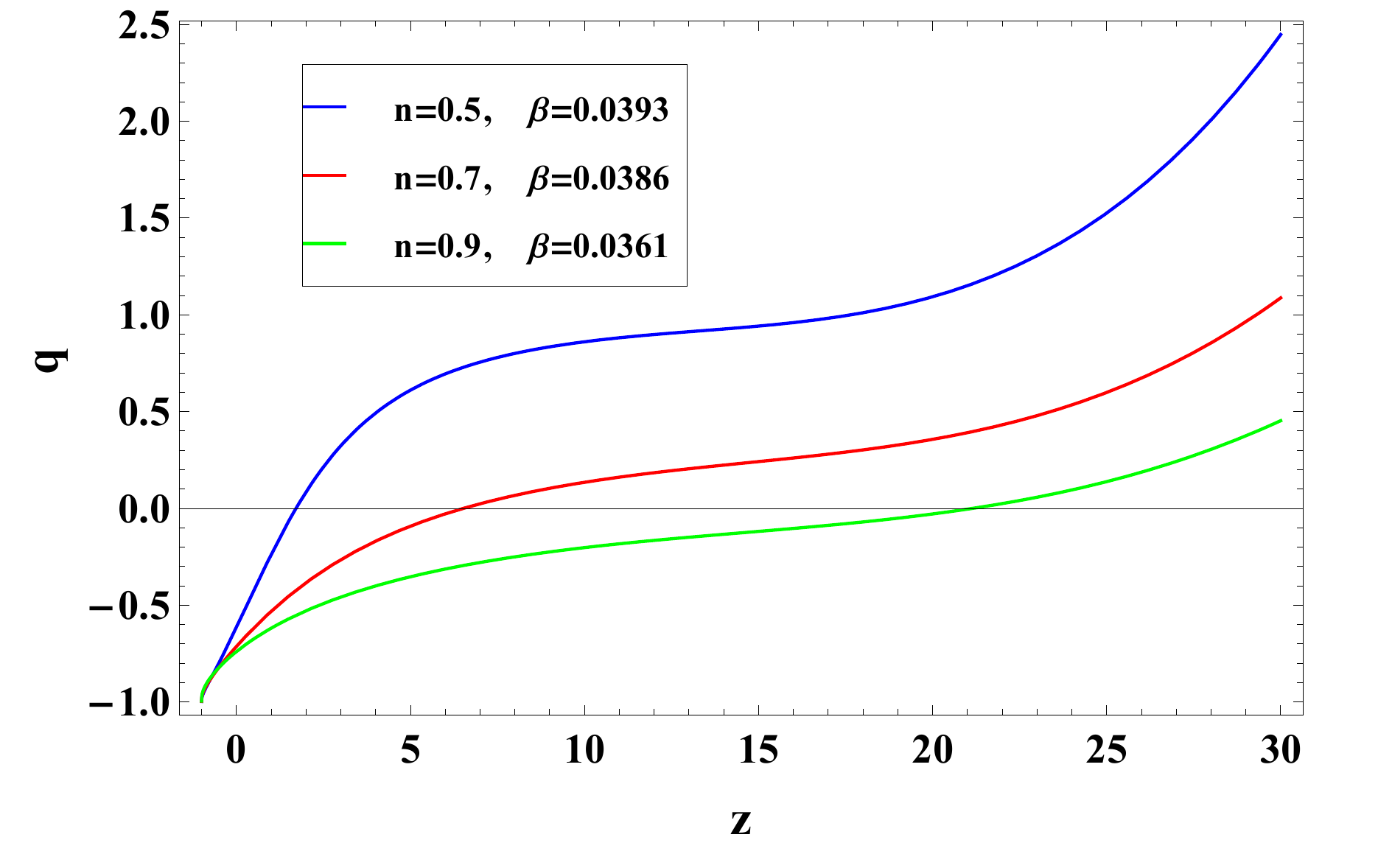}}
	\caption{\scriptsize Depiction of graphs of $ a $, $ H $ and $ q $ vs. $ z $.}
\end{figure}

We considered the following transformation from $ t $ to $ z $ 
\ba \label{16}
&& z= \frac{t_0^n \exp(\beta t_0)}{t^n \exp(\beta t)} -1,
\ea
and calculated the values of the parameters $ a $, $ H $ and $ q $ as functions of $ z $ numerically.  The   visual representations of the scale factor $ a $, the Hubble parameter $ H $ and the deceleration parameter $ q $ as functions of the parameter $ z $ are presented in Fig. 1. 

There is an increase in the values of parameter $ a $ as we move closer to the late Universe and this is depicted in Fig. 1(a). The Hubble parameter $ H $ helps in determining the rate at which the Universe is expanding. The value of $ H $ is positive for all values of redshift $ z $ for $ n=0.5, 0.7, 0.9 $ as shown in Fig. 1(b), and this suggests that the Universe is expanding. For a long period, cosmologists tried to measure the deceleration of expansion caused by gravity. However, the notion of accelerated expansion of the Universe was not considered by anyone, and thus the related parameter was named \textit{Deceleration parameter}. Recent evidences suggest that the Universe's expansion is speeding up. The deceleration parameter $ q $ depicted in  Fig. 1(c) suggest that there has been a phase transition from decelerated expansion to accelerated expansion and Universe will transit to an accelerating de-Sitter regime $ (q = -1) $ for values of $ n $ taken in this model \cite{Bolotin:2015dja}.

\begin{figure}\centering
	\subfloat[]{\label{a}\includegraphics[scale=0.5]{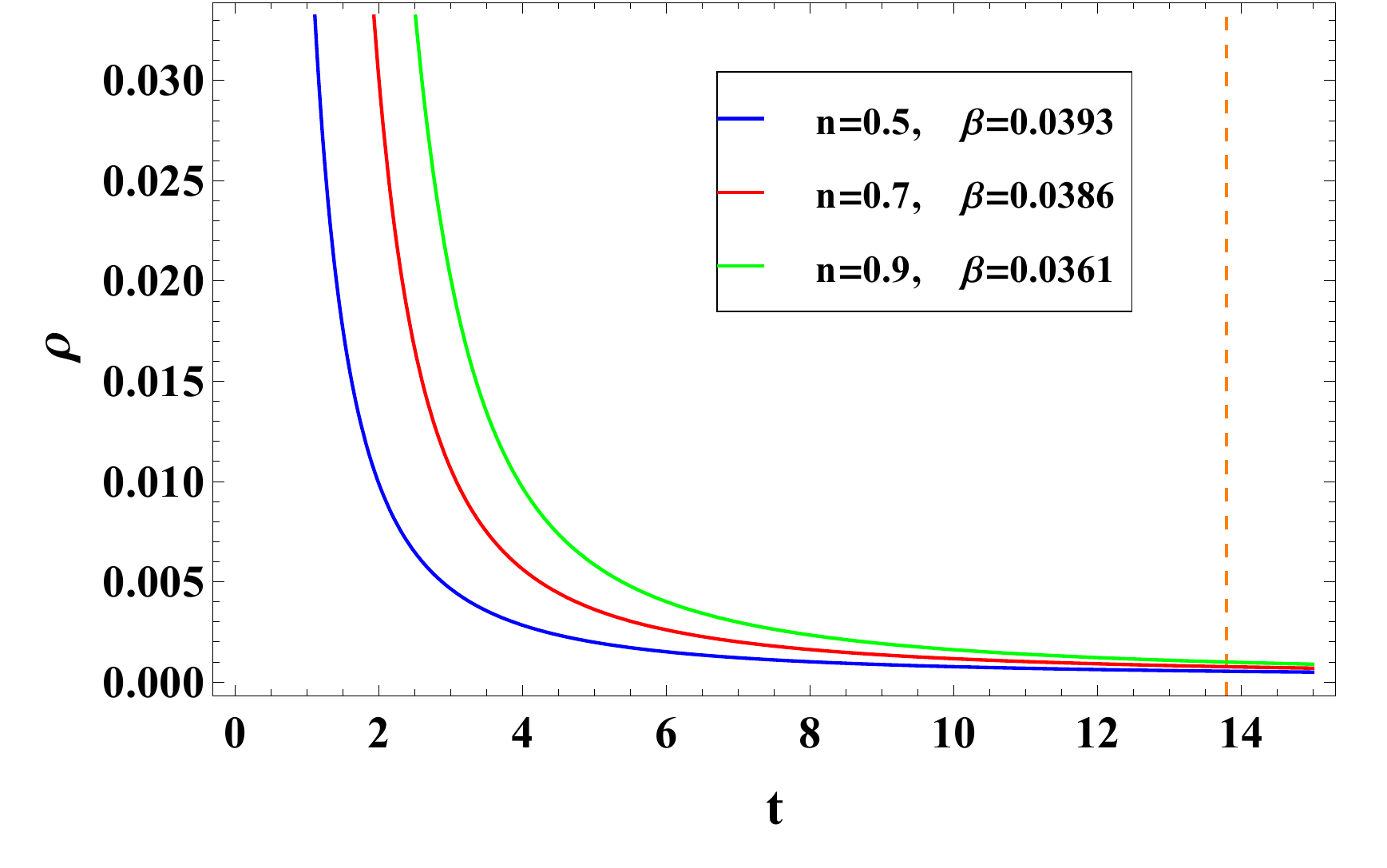}}\hfill
	\subfloat[]{\label{b}\includegraphics[scale=0.4]{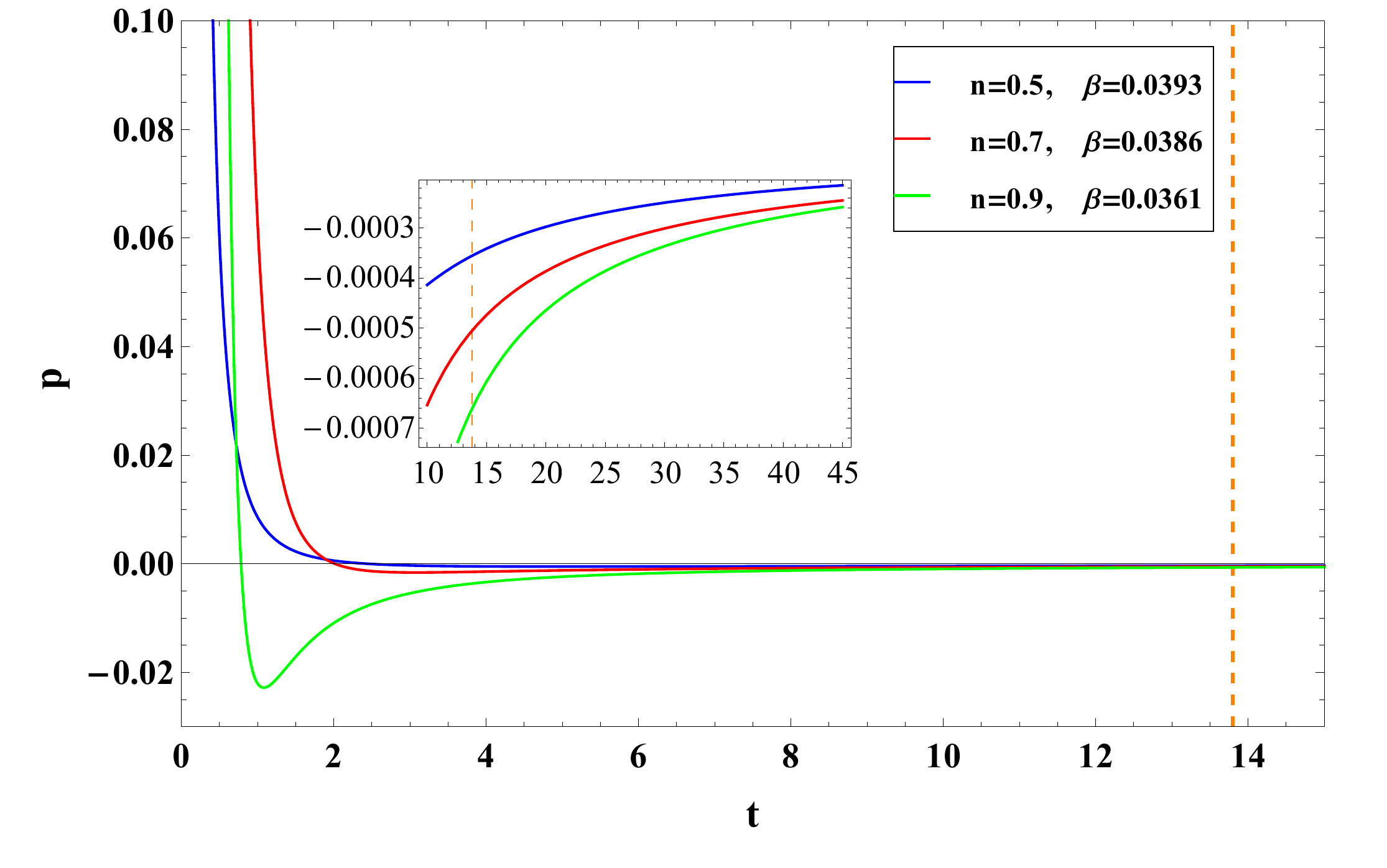}}\par 
	\subfloat[]{\label{c}\includegraphics[scale=0.5]{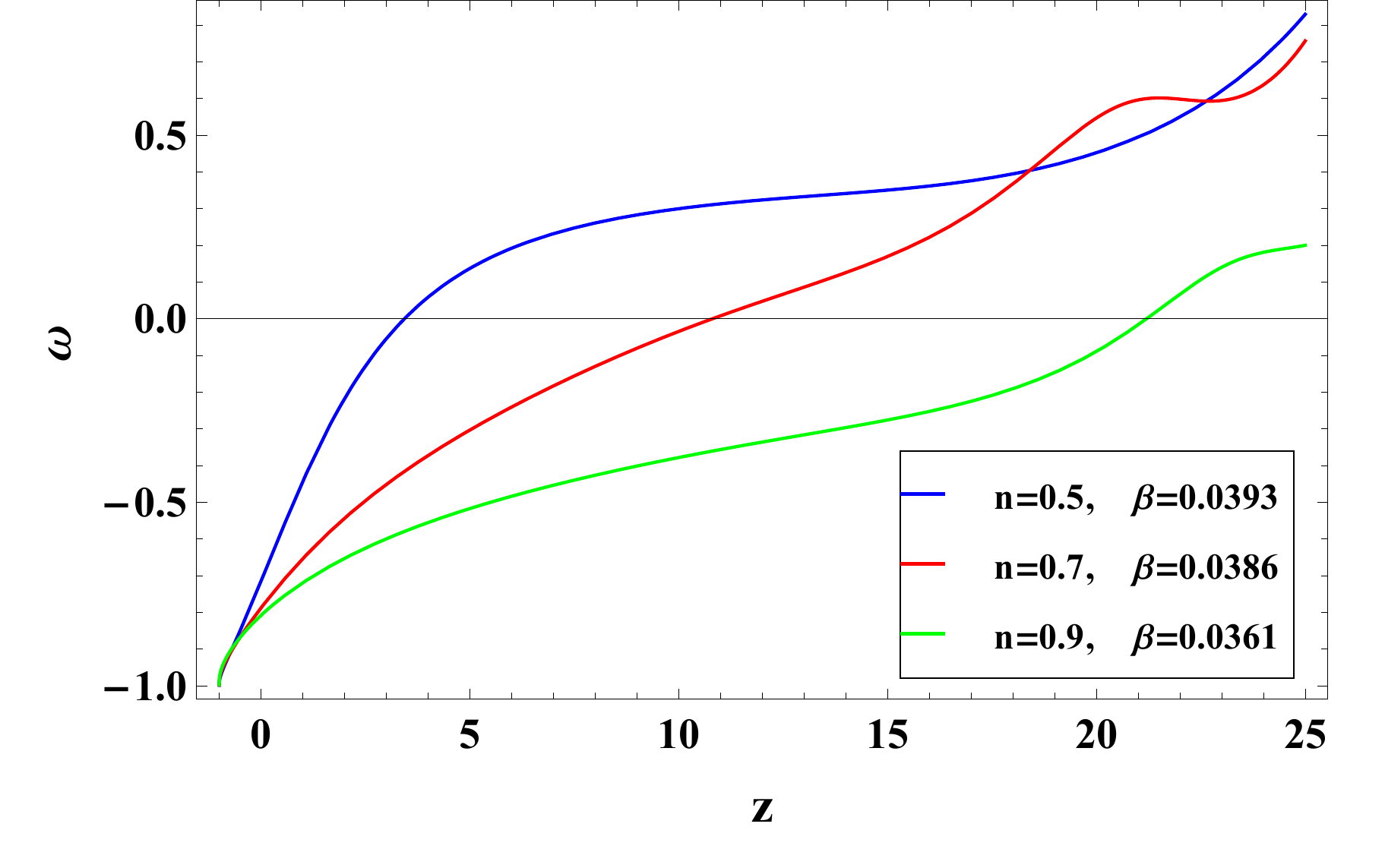}}
	\caption{\scriptsize Depiction of $ \rho-t $, $ p-t $  and $ \omega-z $ plots for $ \alpha=0.5 $ and $ \lambda=2 $.}
\end{figure}

The behaviour of energy density $ \rho $ is displayed in Fig. 2(a) for $ n=0.5, 0.7, 0.9 $. As time progresses from the beginning of the Universe, the value of $ \rho $ decreases, and  approaches $ 0 $ as time $ t $ approaches $ \infty $. For all the values of $ n $ specified earlier, Fig. 2(b) articulates property of pressure of matter for this model. The pressure has a positive value in the early Universe for $ n=0.5, 0.7 $, and started receding with time and eventually approached a negative constant value in late time periods. For $ n=0.9 $, in late times, pressure behaves in a similar way as it does for other values of $ n $ in this model. However in early times, its value dropped from a positive to a large negative value in comparison to $ n=0.5, 0.7 $. According to modern cosmology, negative pressure in the Universe represents acceleration in the cosmos. As a result, the current analysis shows an accelerated phase both now and in the near future.

Using equations in (\ref{11}), the progress of $ \omega =p/\rho $, i.e., the ratio of pressure that dark energy puts on the Universe to the energy per unit volume with respect to the  redshift $ z $ is  depicted in Fig. 2(c) for $ n=0.5, 0.7, 0.9 $.   It is observed that the parameter $ \omega $ is positive in the early Universe and negative in the late Universe with $ \omega=0 $ for each $ n=0.5, 0.7, 0.9 $ and a fixed value of the coupling constant $ \lambda=2 $, for some value of redshift $ z $. The value of $ \omega $ then reaches the quintessence region, and $ \omega \rightarrow -1 $ in late time as $ z \rightarrow -1 $, implying that the Universe's matter initially acts like a perfect fluid. Later on, the model is akin to a dark energy model and exhibits properties like that of a quintessence model and finally approaches $ \Lambda $CDM without entering the phantom zone.

\subsection{Energy conditions}

\qquad The  energy conditions (ECs), which can also be used to generate dynamic model-independent restrictions on the kinematics of the Universe. The components of the energy-momentum tensor are constrained by these criteria, which are based on very general scientific principles. These conditions can be turned into inequalities when picking a model for the medium, limiting the pressure and density values that can be used. We consider the following ECs, which are expressed as the null energy condition (NEC), strong energy condition (SEC), dominant energy condition (DEC), and weak energy condition (WEC) and defined as $ \rho_{eff}+p_{eff} \geq 0 $; $ \rho_{eff}+3p_{eff} \geq 0 $; $ \rho_{eff} > |p_{eff}| \geq 0 $; and $ \rho_{eff} \geq 0 $, $ \rho_{eff}+p_{eff} \geq 0 $ respectively.

From the plots of NEC, SEC and DEC in Fig. 3, we can say that all the energy conditions except SEC hold for $ n=0.5, 0.7, 0.9 $. The violation of SEC in late times signifies the accelerated expansion of the Universe in late times as the inequality $ \rho_{eff}+3p_{eff} \leq 0 $ gives the necessary condition for the accelerated expansion of the Universe \cite{Bolotin:2015dja, Visser:1997qk, Visser:1997tq}.

\begin{figure}\centering
	\subfloat[]{\label{a}\includegraphics[scale=0.40]{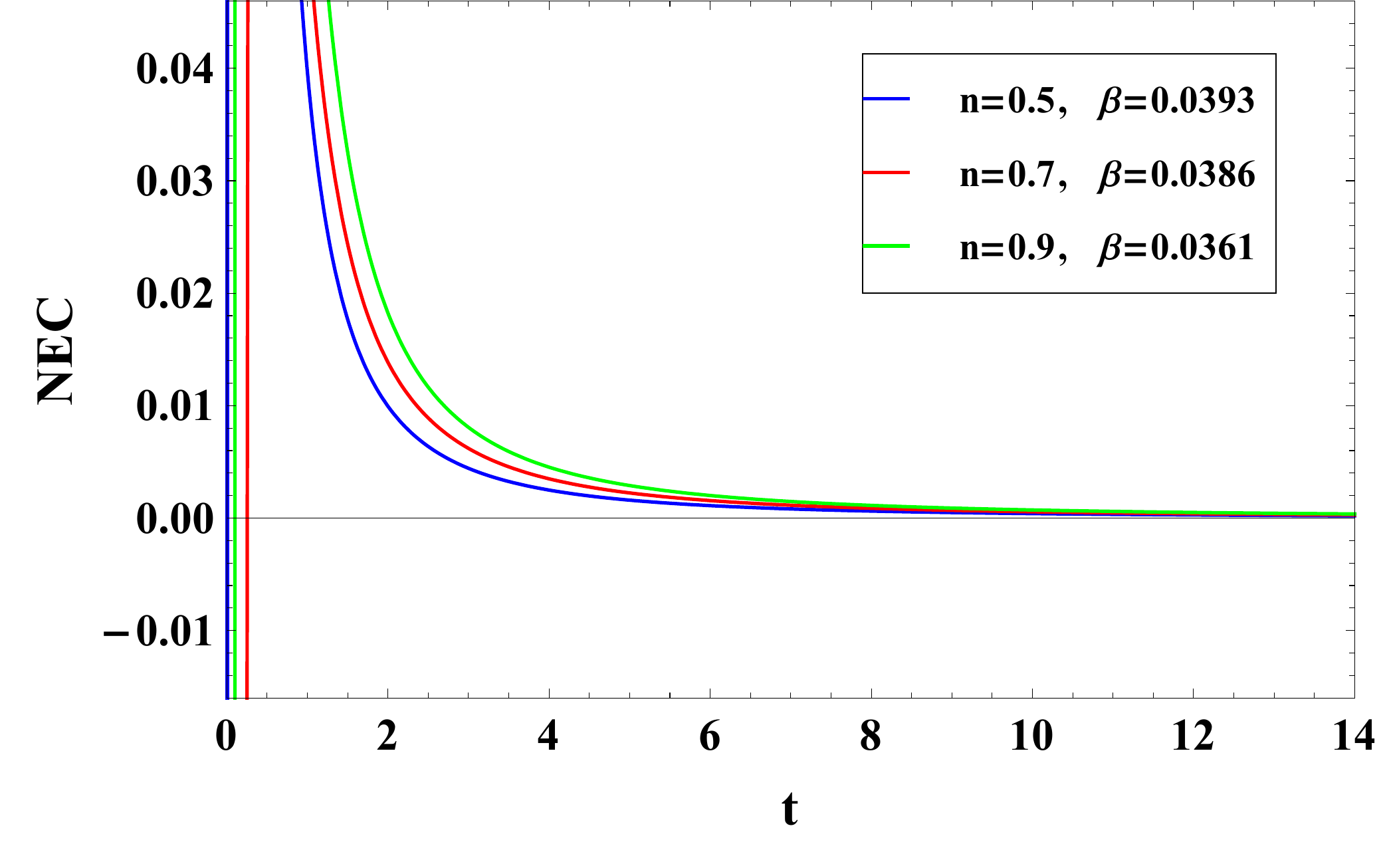}}\hfill
	\subfloat[]{\label{b}\includegraphics[scale=0.40]{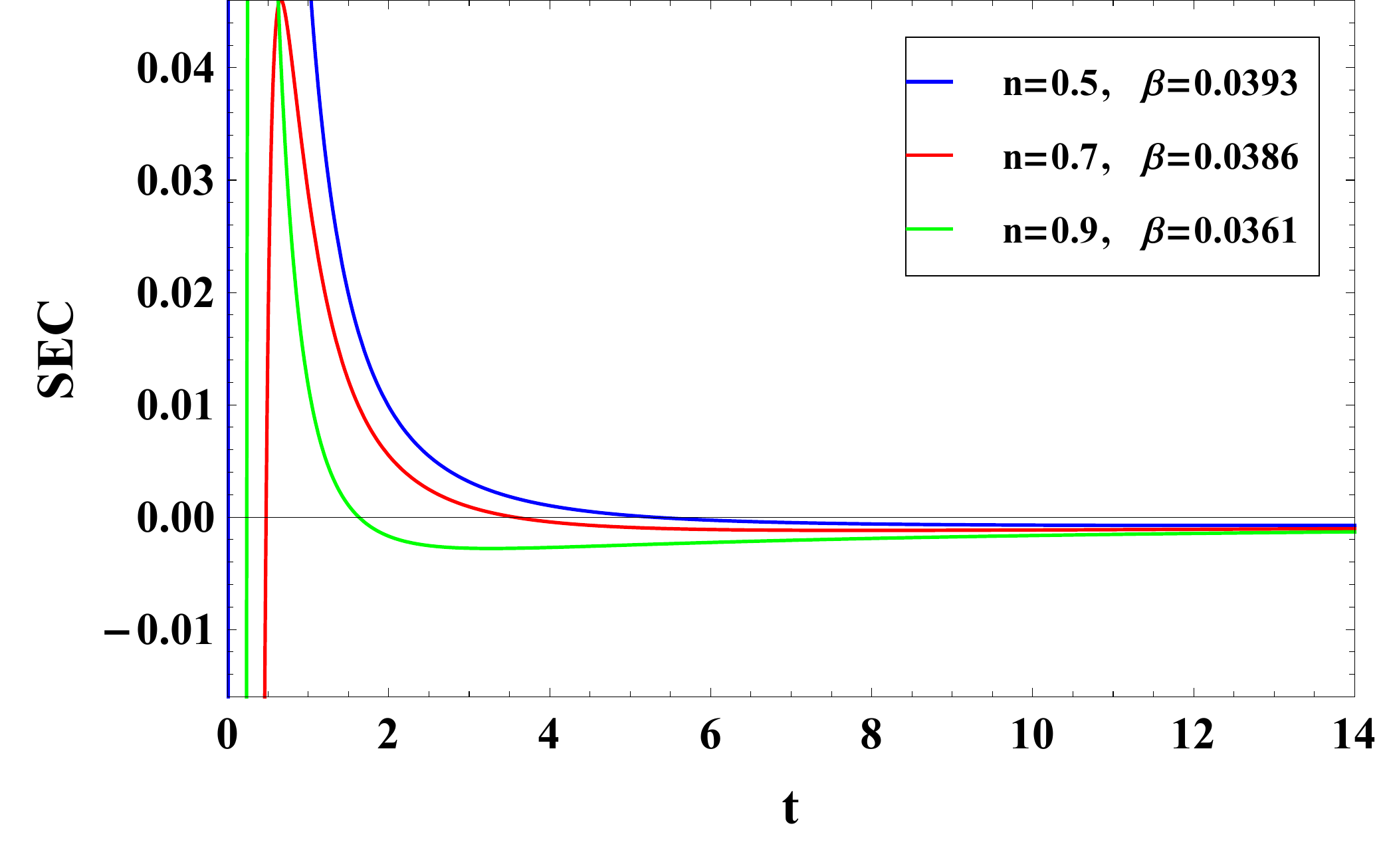}}\par 
	\subfloat[]{\label{c}\includegraphics[scale=0.40]{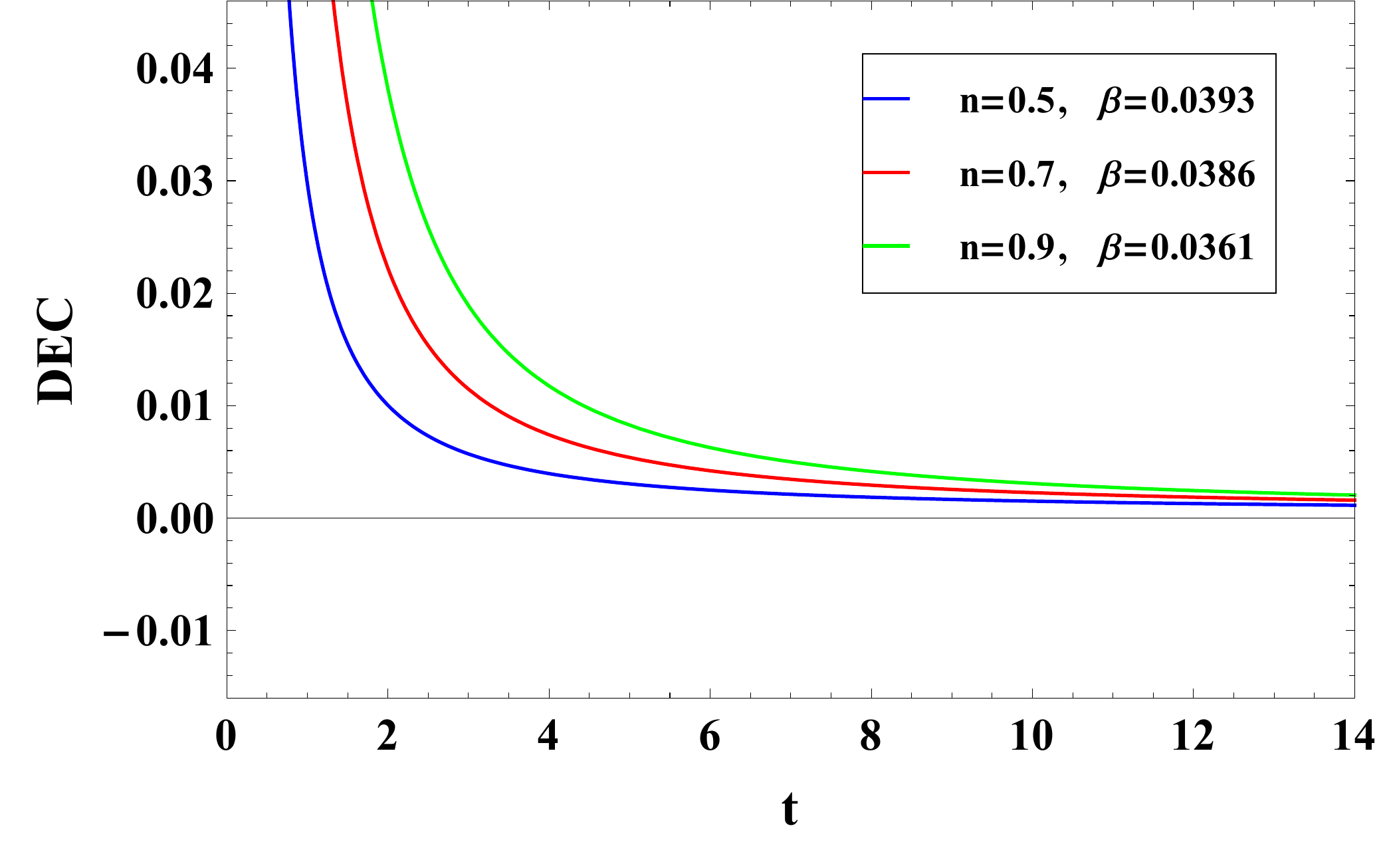}}
	\caption{\scriptsize Depiction of graphs of NEC, SEC and DEC vs. time $ t $ for $ \alpha=-0.01 $.}
\end{figure}

\section{ Dynamics of the model}
The cosmographic analysis of the geometrical parameters is extended by the incorporation of dimensionless higher derivative components of the scale factor $ a(t) $, namely jerk, snap and lerk parameters \cite{Visser:2003vq,Visser:2004bf}. Jerk is also known as jolt at times. Pulse, bounce, impulse, surge, super-acceleration and shock are less commonly used alternative expressions for jerk \cite{Visser:2003vq}. The formula for calculating jerk parameter is given by
\ba \label{17}
&& j=\Big(\frac{d^3a}{dt^3}\Big) \frac{1}{a H^3},
\ea
which on using equations (\ref{13}) and $ (\ref{14})_1 $ can be expressed in terms of $ t $ as  
\ba \label{18}
j=1+\frac{n(2-3n-3\beta t)}{(n+\beta t)^3}.
\ea
Jounce is an another name for snap \cite{Visser:2003vq}. The formula for calculating snap parameter is given by 
\ba \label{19}
&& s=\Big(\frac{d^4a}{dt^4}\Big) \frac{1}{a H^4},
\ea
which on using equations (\ref{13}) and $ (\ref{14})_1 $ can be expressed in terms of $ t $ as 
\ba \label{20}
&& s=1+\frac{3n(n-2)}{(n+\beta t)^4}+\frac{8n}{(n+\beta t)^3}-\frac{6n}{(n+\beta t)^2}.
\ea
Crackle is a term that is occasionally used to refer to the lerk \cite{Visser:2003vq}. The formula for calculating lerk parameter is given by 
\ba \label{21}
&& l=\Big(\frac{d^5a}{dt^5}\Big) \frac{1}{a H^5},
\ea
which on using (\ref{13}) and $ (\ref{14})_1 $, can be expressed in terms of $ t $ as 
\ba \label{22}
&& l=1+\frac{n(4(6-5n)+15(n-2)(n+\beta t)+20(n+\beta t)^2-10(n+\beta t)^3)}{(n+\beta t)^5}.
\ea

The jerk parameter $ j $ is calculated numerically in terms of redshift $ z $ and is presented graphically in Fig. 4(a) for $ n=0.5, 0.7, 0.9 $.  It is observed that the value of $ j $ remains positive for all values of $ n $ in this model and approaches $ 1 $ if $ z $ approaches $ -1 $. This finding is compatible with typical $ \Lambda $CDM observations, although at the moment $ z = 0 $, $ j \neq 1 $ for all $ n $. As a result, for every $ n $, our model is comparable to the dark energy model, and differs from $ \Lambda $CDM.\\

\begin{figure}\centering
	\subfloat[]{\label{a}\includegraphics[scale=0.5]{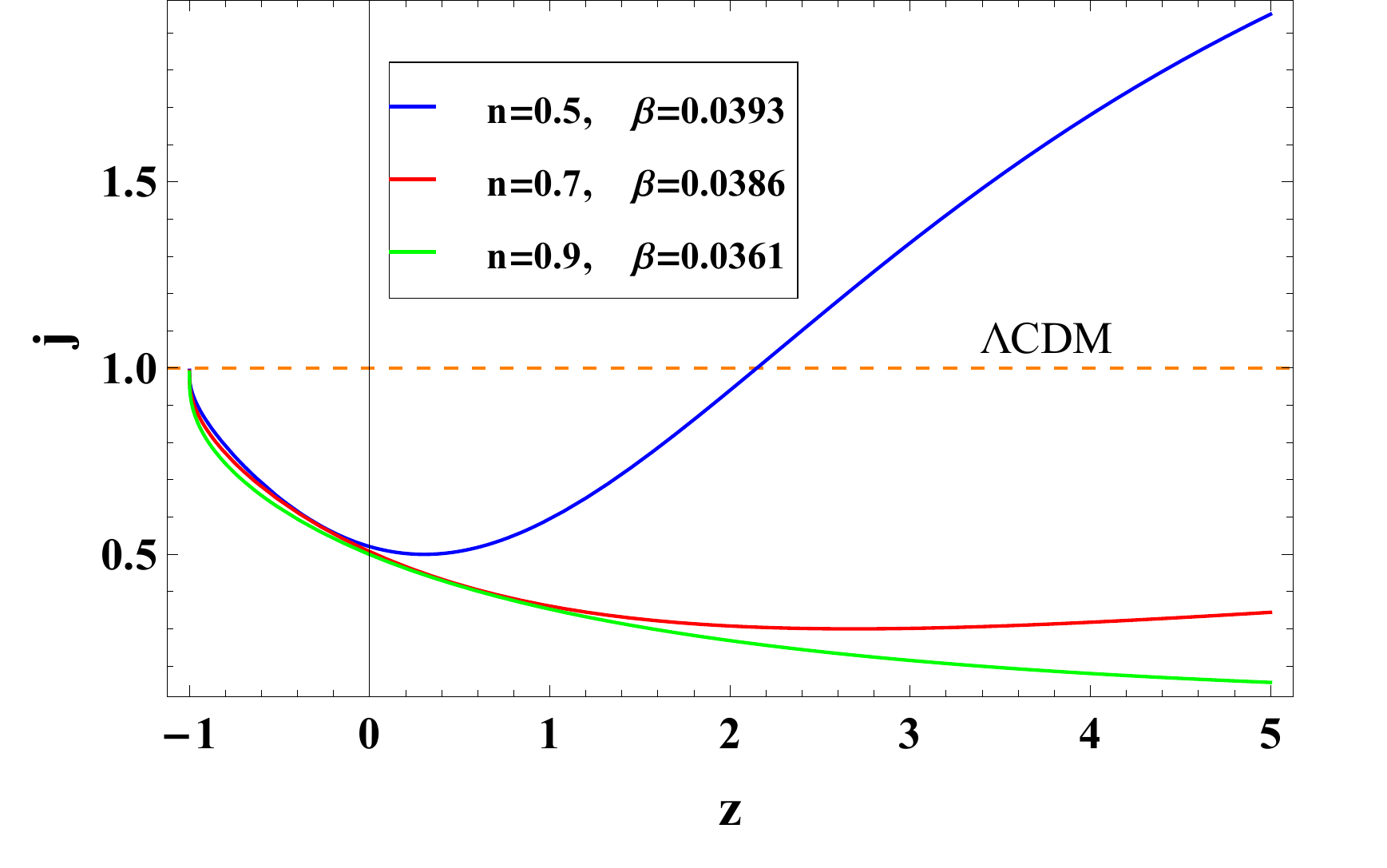}}\hfill
	\subfloat[]{\label{b}\includegraphics[scale=0.5]{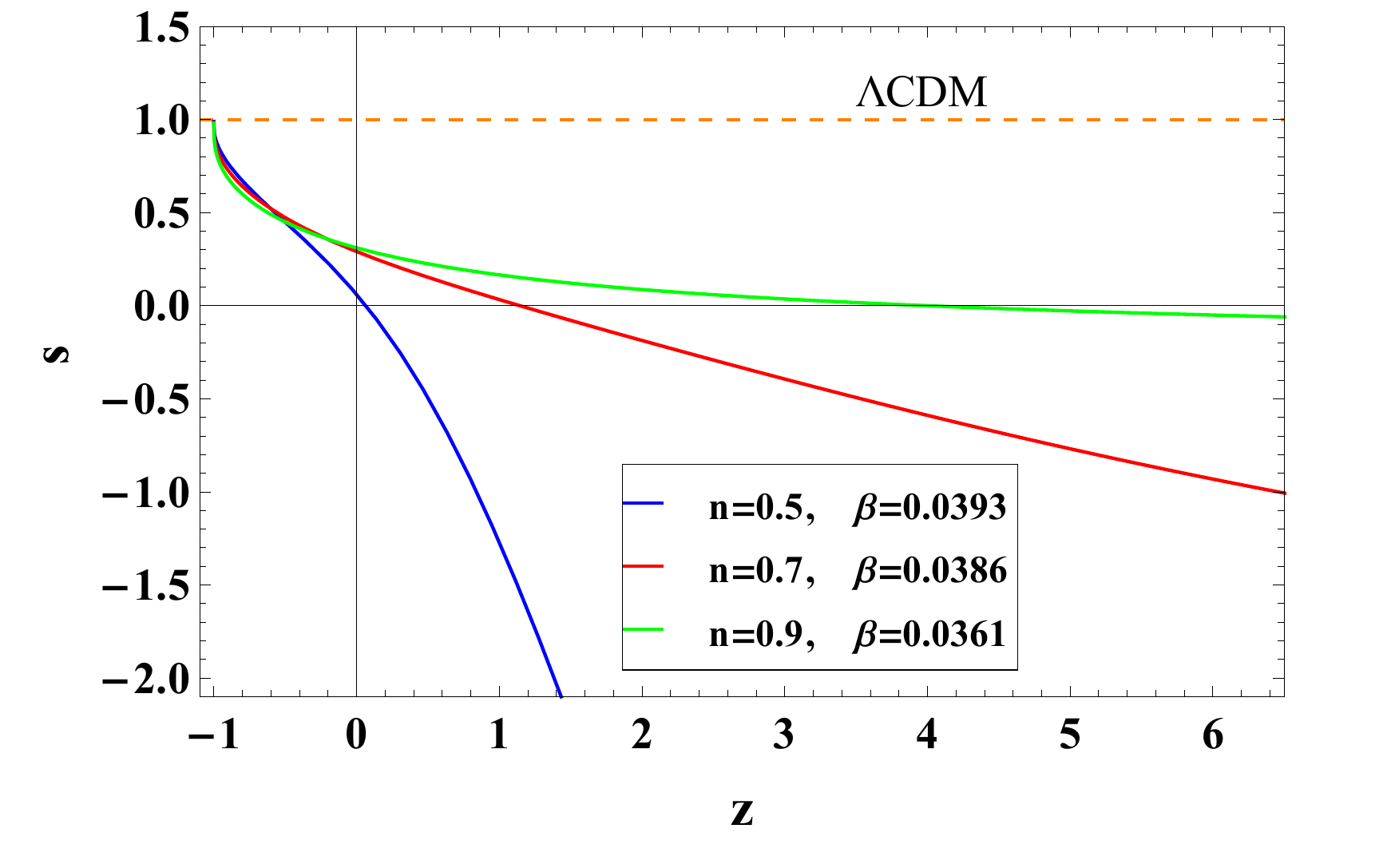}}\par 
	\subfloat[]{\label{c}\includegraphics[scale=0.5]{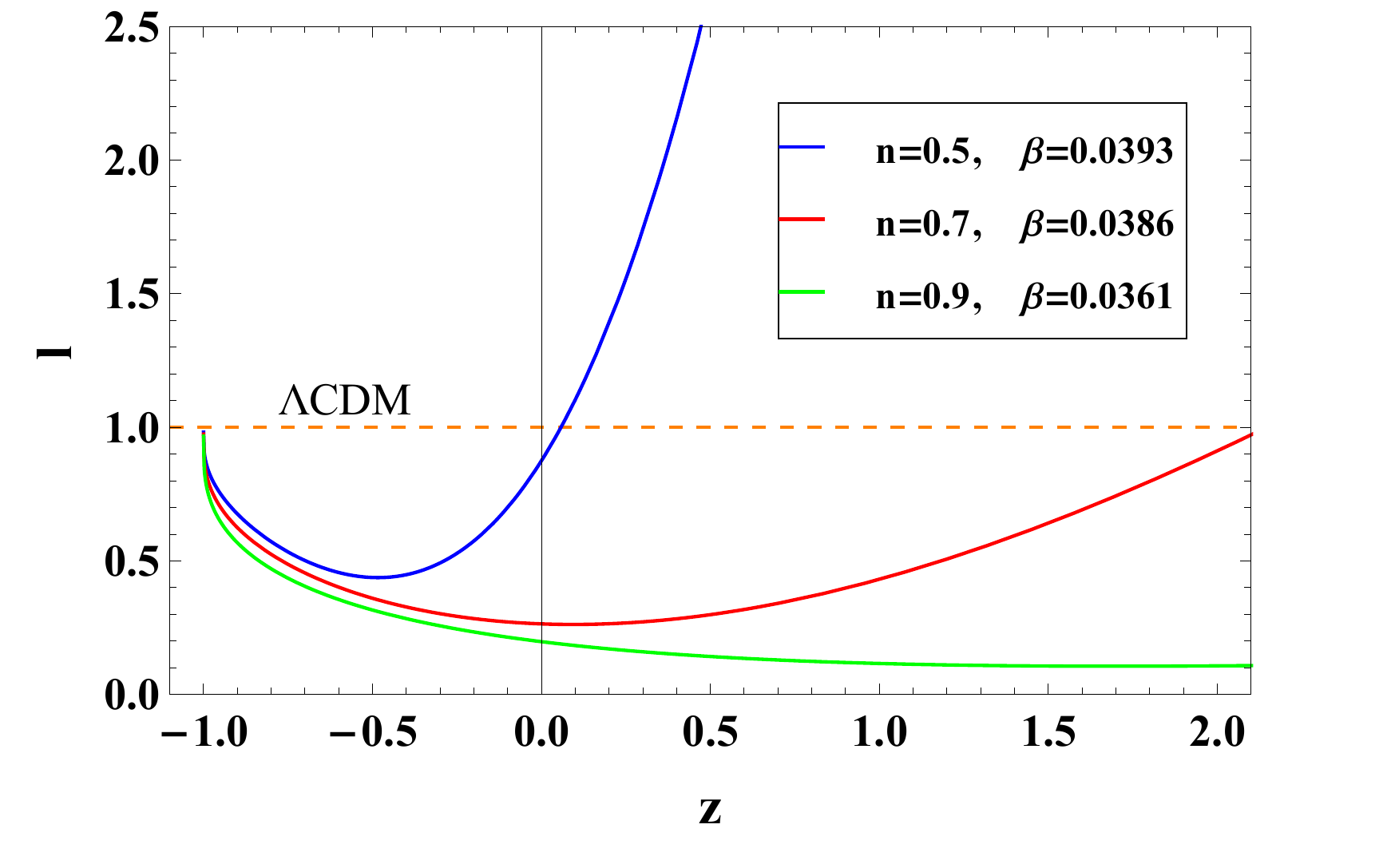}}
	\caption{\scriptsize Depiction of graphs of Jerk parameter $ j $ , Snap parameter $ s $ and Lerk parameter $ l $ vs. $ z $.}
\end{figure}

The snap $ s $ and lerk $ l $ parameters are calculated numerically in terms of redshift $ z $ and are depicted  in Fig. 4(b) and Fig. 4(c), respectively for $ n=0.5, 0.7, 0.9 $. The profile of snap parameter $ s $ during its evolution indicates that, it takes values in the negative range for all $ n $ in the early Universe, then takes values in the positive range as the Universe evolves, i.e., there is just one transition from negative to positive range throughout the complete evolution of $ s $. Also as it is evident, the transition of $ s $ is dependent on the model parameter $ n $, i.e., the transition of $ s $ occurs early as $ n $ varies between $ 0.5 $ and $ 0.9 $. Fig. 4(c) shows the variation of the lerk parameter $ l $ with respect to the redshift $ z $. Without any redshift transition, the lerk parameter $ l $ takes up only positive values. On the similar line of jerk parameter $ j $, both snap parameter $ s $ and lerk parameter $ l $ approach 1 in late time i.e., for $ z \rightarrow -1 $.

\subsection{ Om diagnostic}

Om diagnostic can be conveniently used to distinguish between standard $ \Lambda $CDM model and other dark energy models \cite{Sahni:2008xx,Zunckel:2008ti}. $ Om(z) $ is used to denote Om diagnostic and defined as
\ba \label{23}
&& Om(z)=\frac{\Big(\frac{H(z)}{H_0}\Big)^2-1}{z(z^2+3z+3)}.
\ea
The diverse \textit{Om(z)} pathways allow for major differences between multiple dark energy models. Fig. 5(a) illustrates quintessence type behaviour $ (0 > \omega > -1) $ as the trajectories of $ Om(z) $ show convexity \textit{w.r.t.} to the z-axis and also display a declining trend as redshift $ z $ increases for $ n=0.5, 0.7, 0.9 $.\\

\begin{figure}\centering
	\subfloat[]{\label{a}\includegraphics[scale=0.5]{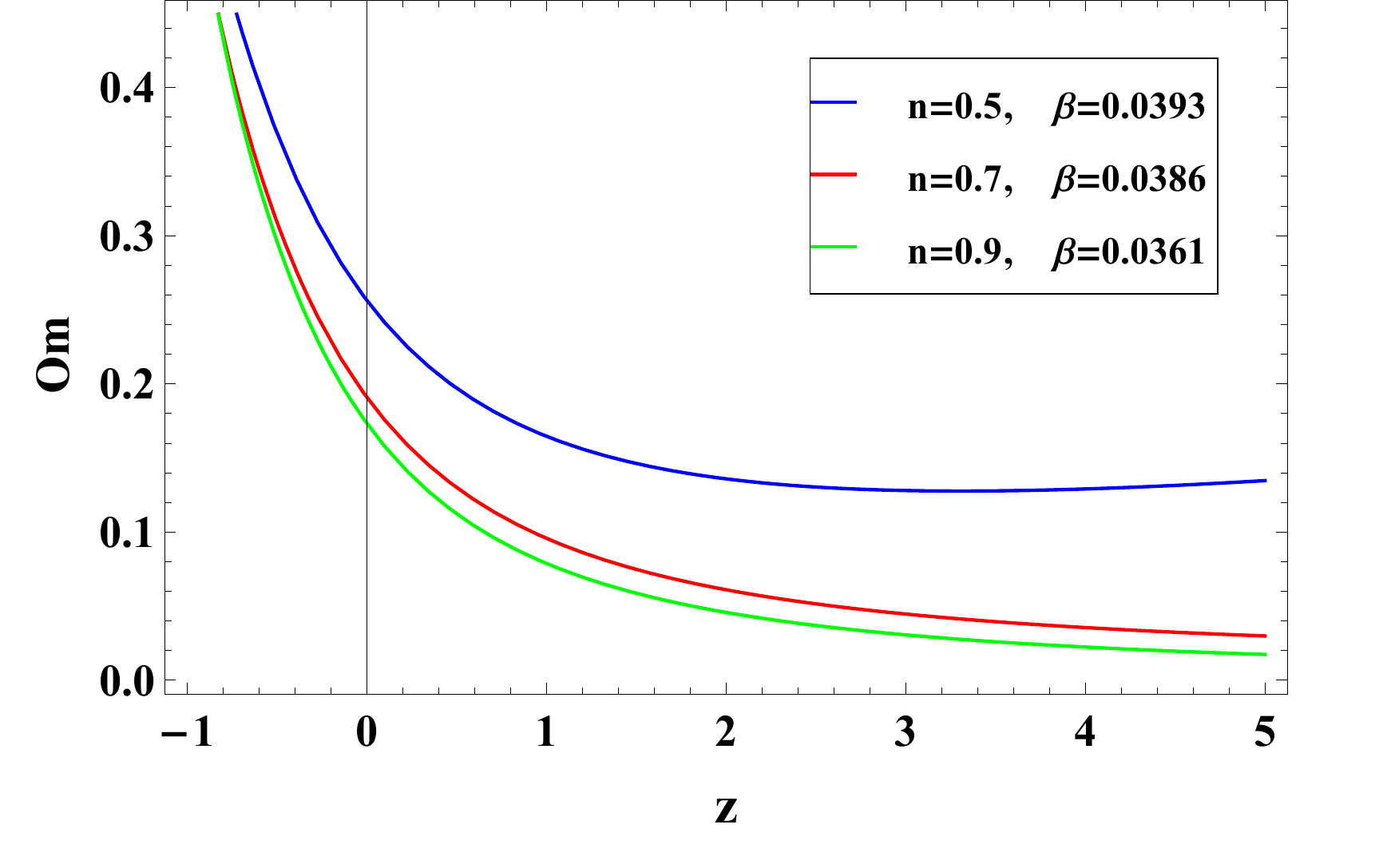}}
	\subfloat[]{\label{b}\includegraphics[scale=0.5]{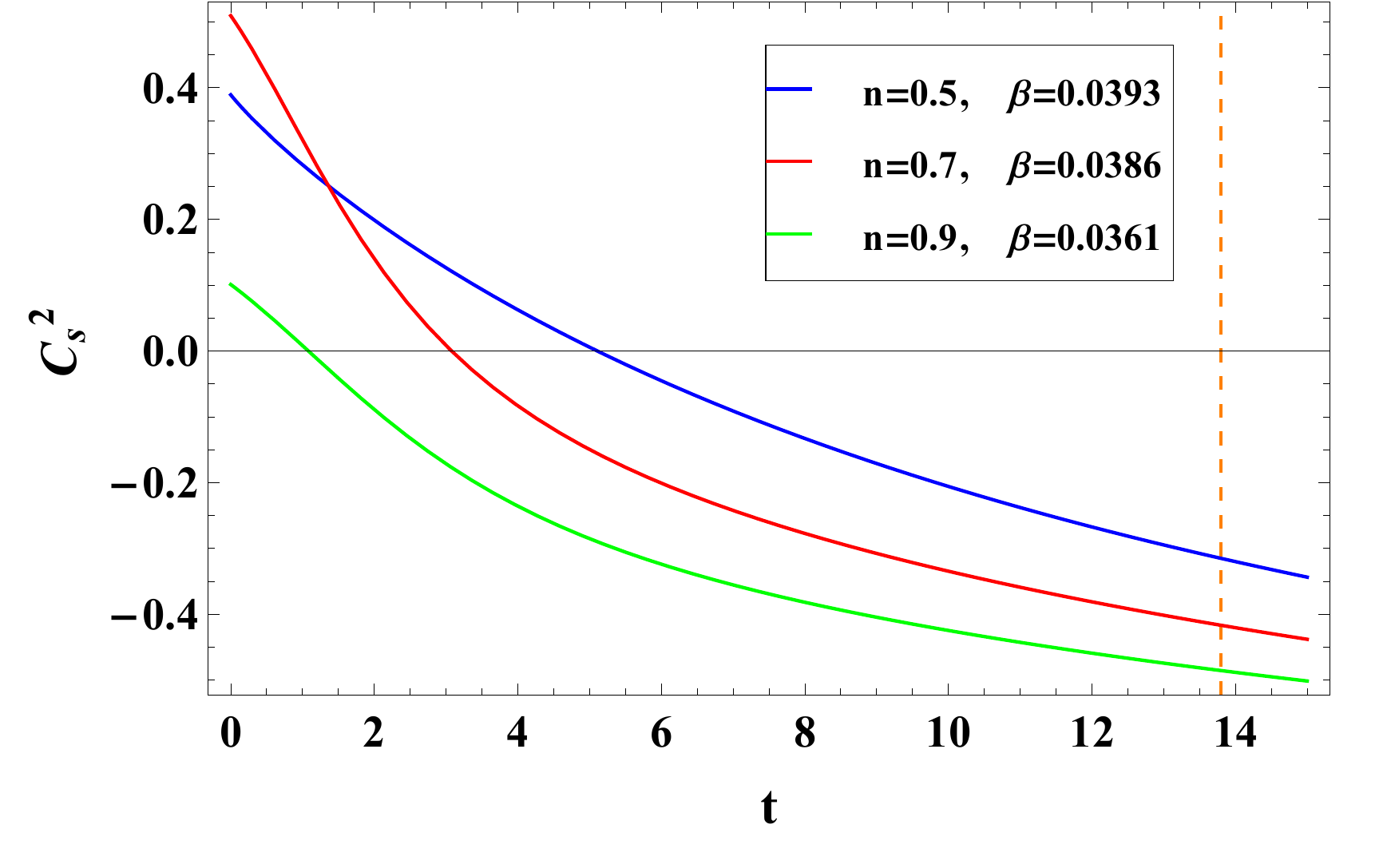}} 
	\caption{\scriptsize Depiction of graphs of (a) Om diagnostic $ Om $ vs. redshift $ z $, (b) Velocity of sound $ C_s^2 $ vs. time $ t $.}
\end{figure}

\subsection{ Velocity of sound}

Velocity of sound $ C_{s}^{2} $ is defined as
\ba \label{24}
&& C_s^2=\frac{dp}{d\rho}.
\ea
If a model satisfies the condition $ 0 \leq C_{s}^{2} \leq 1 $ then that model will be stable. From the graphical representation of the velocity of sound versus time $ t $ in Fig. 5(b), it can be inferred that for $ n=0.5 $ the model will be stable for $ t $ up to $ 6 $ approximately, for $ n=0.7 $ the stability of the model will be up to $ t=4 $ approximately,  and  for $ n=0.9 $ the stability will be up to $ t=1 $ approximately. The degree of unstability of this model is very high in late time. 

\subsection{ Statefinder diagnostic}

The Statefinder diagnostic is a geometrical analysis to describe the phenomena of distinct DE models by the parameters $ r $, and $ s^* $, which are defined as
\ba \label{25}
r=\frac{\dddot{a}}{aH^{3}}\text{, \ \ }s^*=\frac{r-1}{3(q-\frac{1}{2})},
\ea
where $ q\neq \frac{1}{2} $. The sequence of parameters $ \lbrace s^*,r \rbrace$ was proposed by Sahni et al. \cite{Sahni:2002fz,Sahni:2002yq} and Alam et al. \cite{Alam:2003sc}, where the variable represented by $ r $ is identical to the jerk parameter $ j $. A specific linear combination of the jerk (or can also say $ r $) and deceleration parameters $ q $ is used as another ``statefinder'' variable represented by $ s^* $. For the value of scale factor $ a $ in equation (\ref{13}), the parameters $ r $ and $ s^* $ can be expressed in terms of $ t $ as
\ba \label{26}
&& r=1+\frac{n(2-3n-3\beta t)}{(n+\beta t)^3}, \htwo 
s^*=\frac{n(2-3n-3 \beta t)}{3(n+\beta t)^3  \Big(\frac{-3}{2}+\frac{n}{(n+\beta t)^2}\Big)}.
\ea

\begin{figure}\centering
	\subfloat[]{\label{a}\includegraphics[scale=0.35]{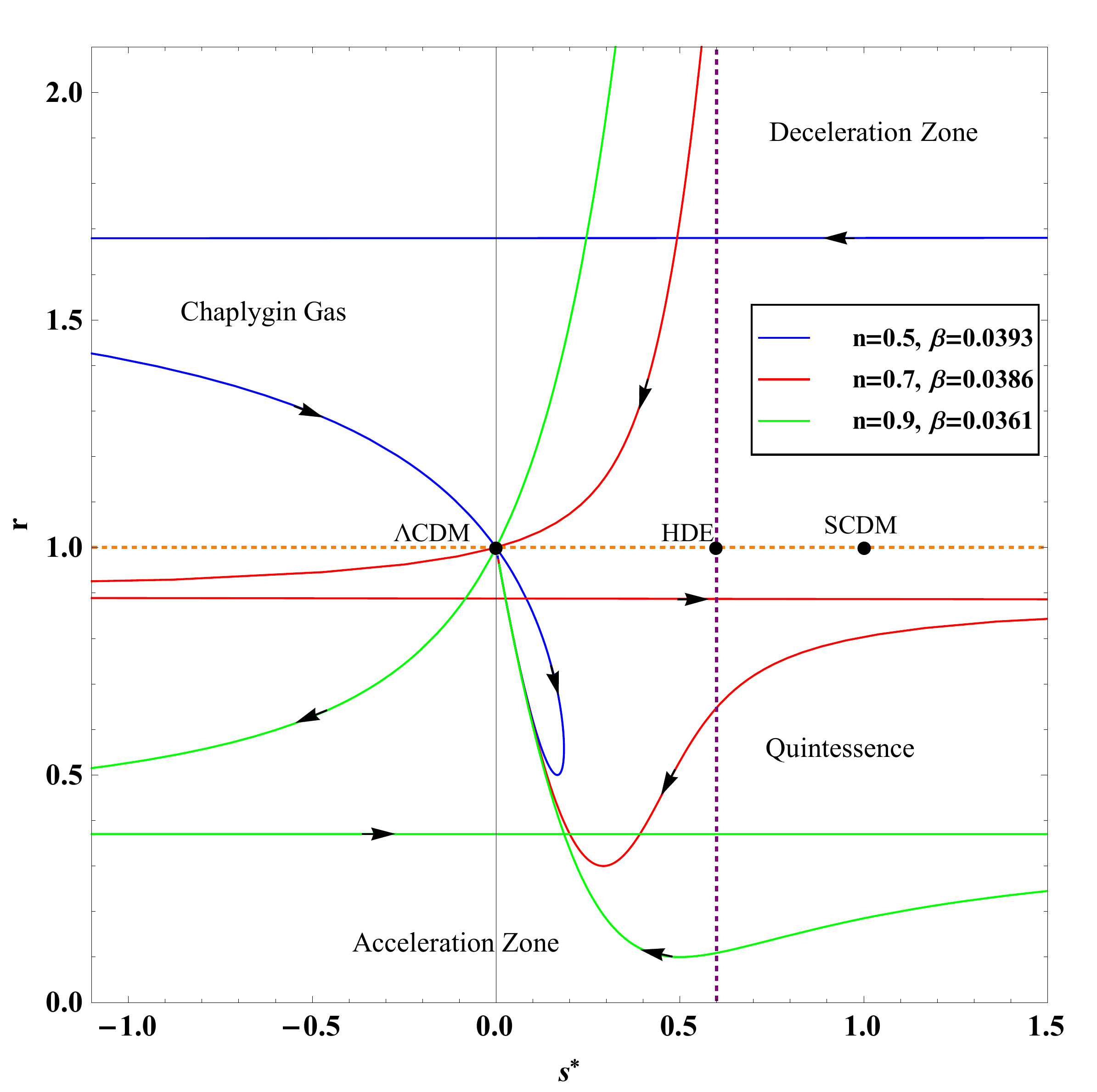}}\hfill
	\subfloat[]{\label{b}\includegraphics[scale=0.40]{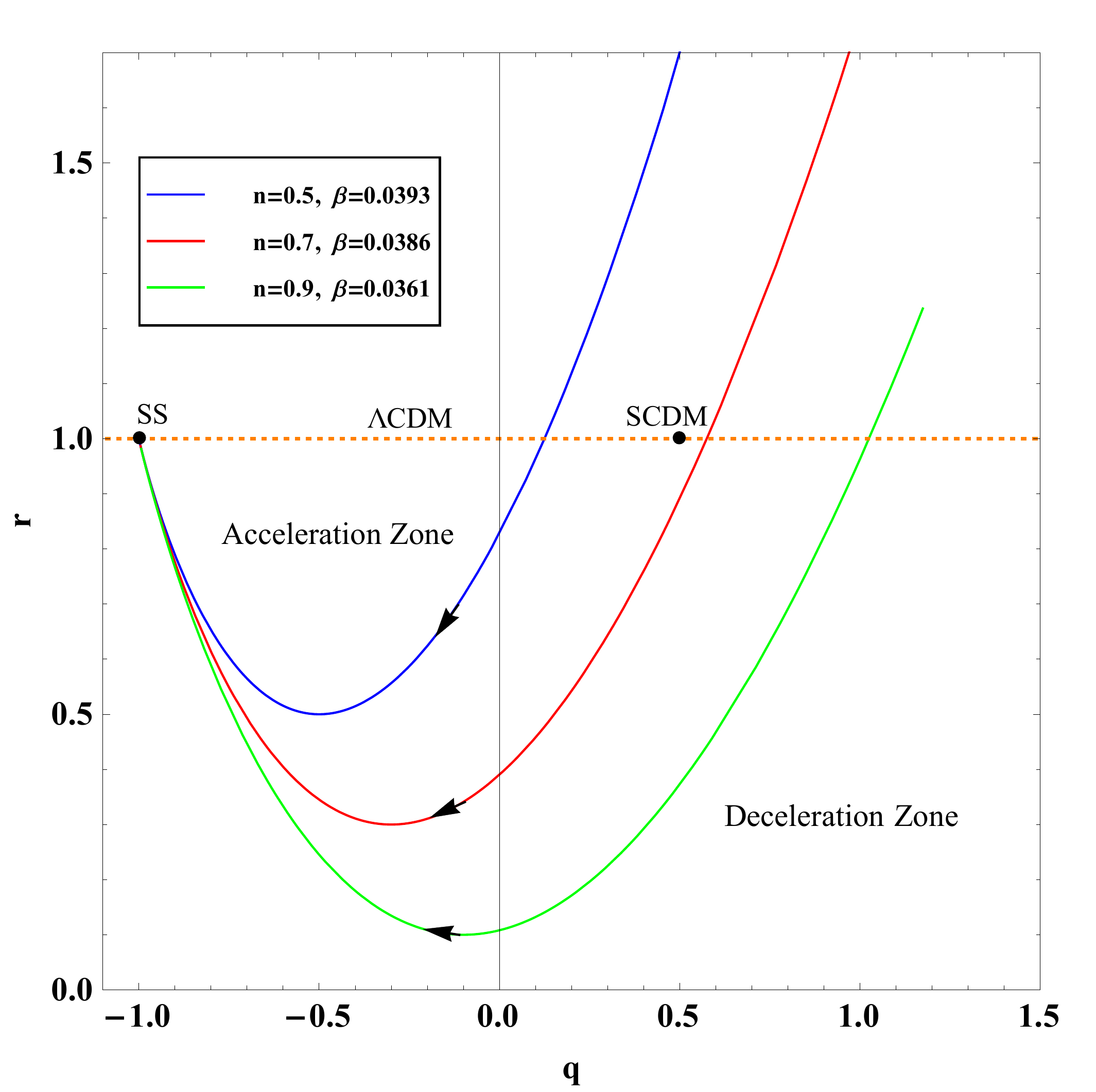}} 
	\caption{\scriptsize Depiction of $ s^*-r $ and $ q-r $ trajectories.}
\end{figure}

The values of $ r $ and $ s^* $ calculated numerically are plotted in Fig. 6(a) as the three time-dependent trajectories of this model in the $ s^*-r $ plane for all $ n $. Arrows represent the directions of the $ s^*-r $ trajectories in the figure. Each trajectory for $ n=0.5, 0.7, $ and $ 0.9 $ follows the same pattern, beginning in the region $ r>1, s^*>0 $ and progressing through the quintessence region ($ r<1, s^*>0 $) and approaches the $ \Lambda $CDM ($ r=1,s^*=0 $). The $ n=0.5 $ trajectory also travels through Chaplygin gas ($ r>1,s^*<0 $). All of the trajectories deviate from $ SCDM $, where $ SCDM $ ($ r=1,s^*=1 $) is similar to a matter-dominated Universe. The holographic dark energy ($ HDE $) paradigm is represented by the point with coordinates $ r=1, s^*=\frac{2}{3} $ on the horizontal line.

In the $ q-r $ plane, Fig. 6(b) depicts the three trajectories with time for all $ n $ in this model. It may be noted that $ q $ changes its sign from positive to negative, which is already observed  in Fig. 1(c). The trajectory of each value of $ n $ continues in the vicinity of $ SCDM $ ($ r=1,q=1/2 $) but never converges to $ SCDM $. These  trajectories  cross the $ \Lambda $CDM line, which is the line $ r=1 $ parallel to the $ q $-axis in $ q-r $ plane. As time passes, the values of $ r $ and $ q $ begin to decrease until they reach their lowest point, following which they begin to rise towards the point marked as $ SS $ ($ r=1,q=-1 $), the Universe's steady state model. Our dark energy model may behave like the steady state model in late times, based on the evolution of the trajectories to $ SS $.

\section{ Some other cosmological tests}
In the past $ t_L $ was the temporal gap between the production of light from the source and the reception of light on the Earth. As a result, the total time $ t_L $ elapsed between the galaxy's light beam emitting at time $ t_z $ for a specific redshift $ z $ and reaching us at a time $ t $ for redshift $ z = 0 $ is represented as \cite{Nagpal:2019vre}
\ba \label{27}
t_L= t_0-t_z=\int_{a}^{a_0} \frac{dt}{\dot{a}},
\ea
where, the value of $ t_z $ can be calculated numerically.  Astronomers might benefit from this technique. The farther away an object is, the further back in time we are witnessing its light. Look-back time is what allows us to understand the evolution of galaxies through time by observing them from varying distances.  The variations of look-back time $ t_L $ with respect to the redshift $ z $ is presented in Fig. 7(a) and it is observed that the values decrease as we move closer to the late Universe.

\begin{figure}\centering
	\subfloat[]{\label{a}\includegraphics[scale=0.45]{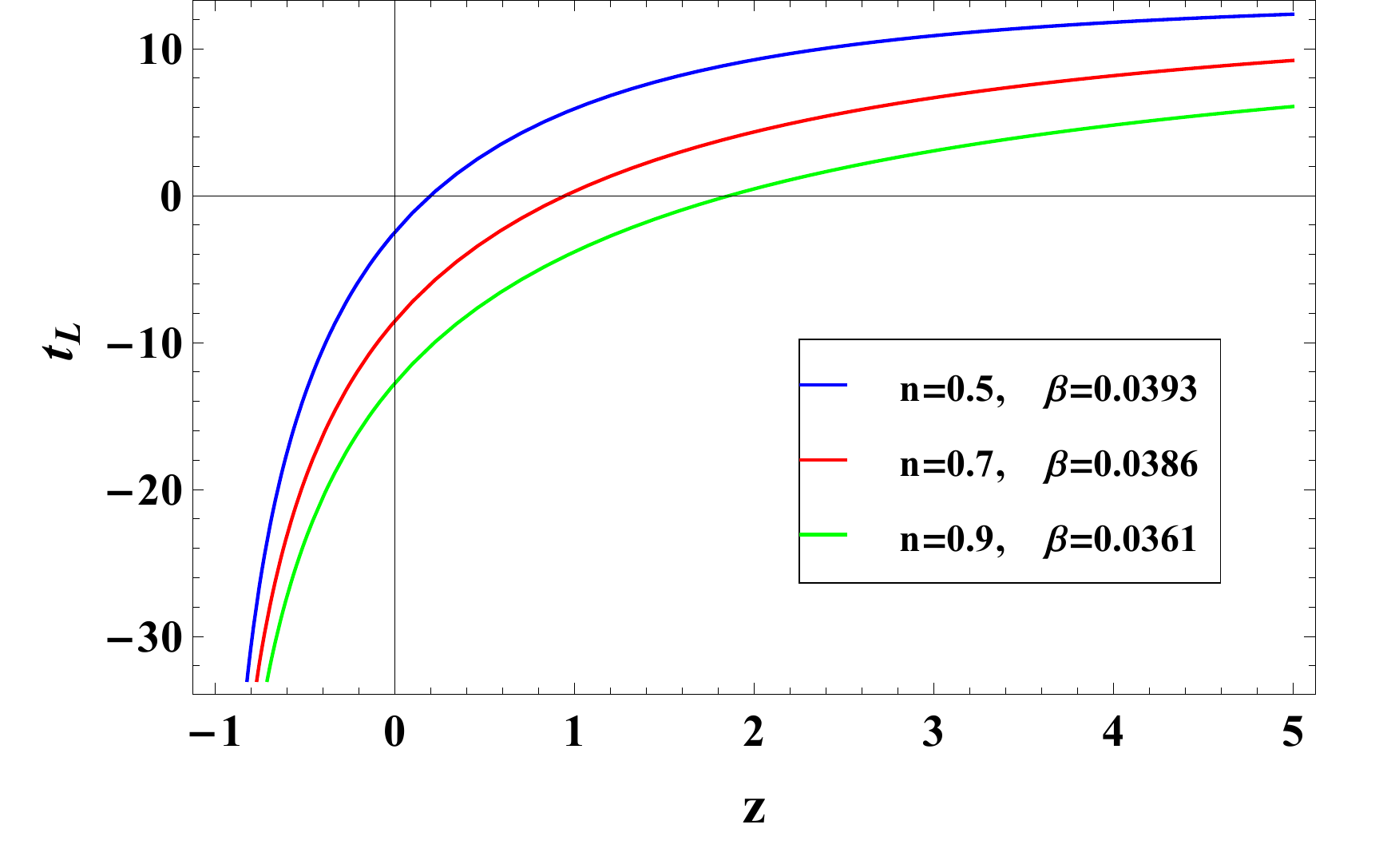}}
	\subfloat[]{\label{b}\includegraphics[scale=0.45]{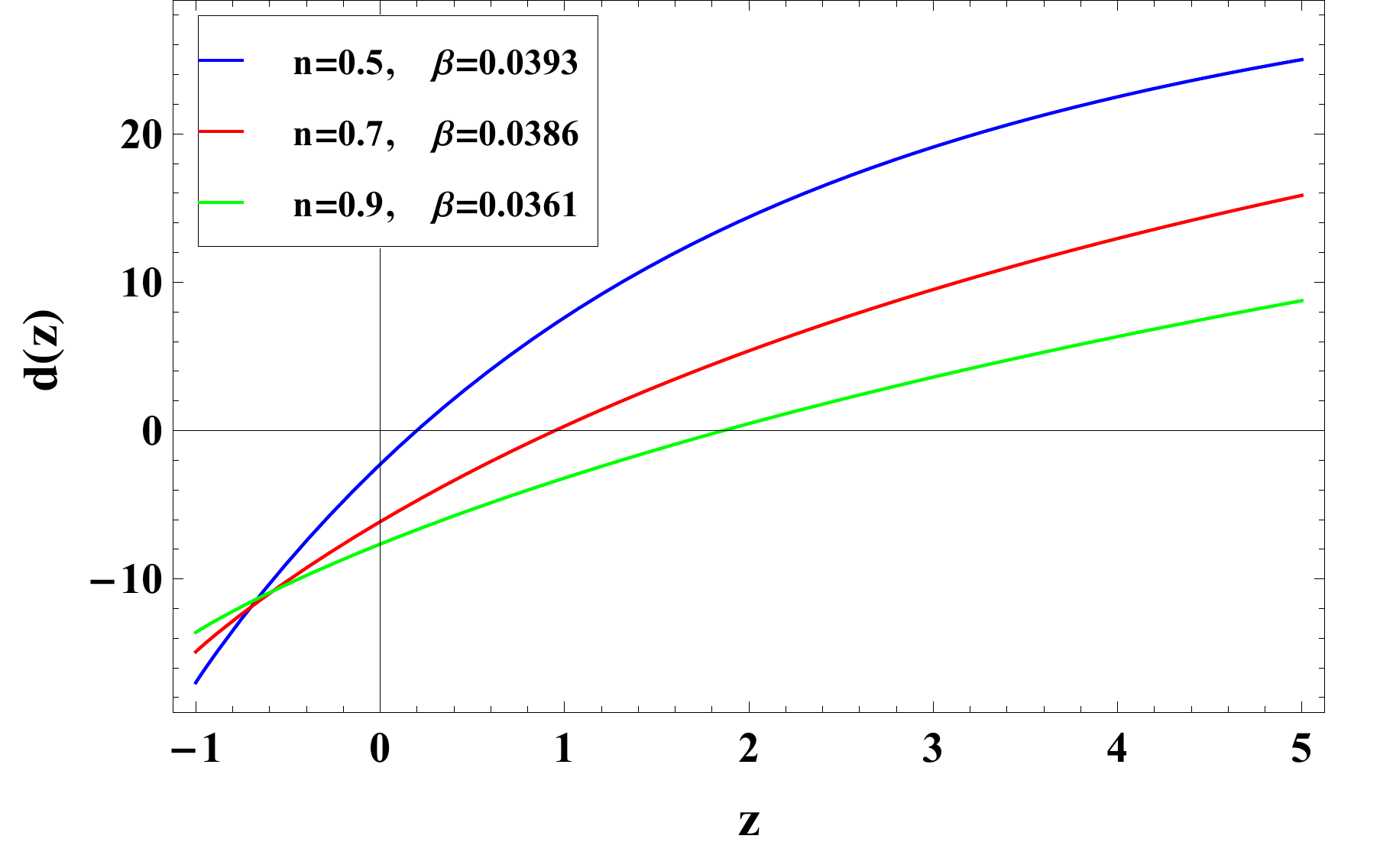}} 
	\caption{\scriptsize Depiction of graphs $ t_{L} $ and $ d(z) $ vs. $ z $.}
\end{figure}

\subsection{ Proper distance}

The instantaneous distance between source and observer at the time of detection is known as proper distance. It  is formulated as $ d(z)=a_0 r(z) $, where $ r(z) $ is the radial distance obtained by
\ba \label{28}
&& r(z)=\int_t^{t_0} \frac{dt}{a(t)}\,\,.
\ea
Using the equation (\ref{13}) in (\ref{28}), we obtain
\ba \label{29}
&& r(z)=t^{(1-n)} E_n(\beta t) - \beta ^{(-1+n)} \Gamma(1-n,13.8 \beta), 
\ea
where $ E_n $ is the exponential integral funtion and is defined as $ \disp E_n(\alpha) = \int_1^{\infty} ({e^{-\alpha t}}/{t^n}) ~dt $. From (\ref{29}), we obtain the following form of $ d(z) $ 
\ba \label{30}
&& d(z)=a_0 (t^{(1-n)}E_n(\beta t) -  \beta ^{(-1+n)} \Gamma(1-n,13.8 \beta)).
\ea
The variations of the proper distance $ d_z $ with respect to the redshift $ z $ is presented in Fig. 7(b) and it is observed that the values of $ d_z $ decrease as we move closer to the late Universe.

\subsection{ Angular diameter}

The angular diameter $ d_A $ as a function of $ z $ is formulated as
\ba \label{31}
&& d_A= \frac{d(z)}{1+z}.
\ea
Using equation (\ref{30}) in (\ref{31}) we obtain
\ba \label{32}
&& d_A=\frac{a_0 (t^{(1-n)} E_n( \beta t) - \beta ^{(-1+n)} \Gamma(1-n,13.8 \beta))}{1+z}.
\ea
The variations of  angular diameter  $ d_A $ with respect to the redshift $ z $ is presented in Fig. 8(a) and it is observed that the values decrease as we move closer to the late Universe.

\begin{figure}\centering
	\subfloat[]{\label{a}\includegraphics[scale=0.45]{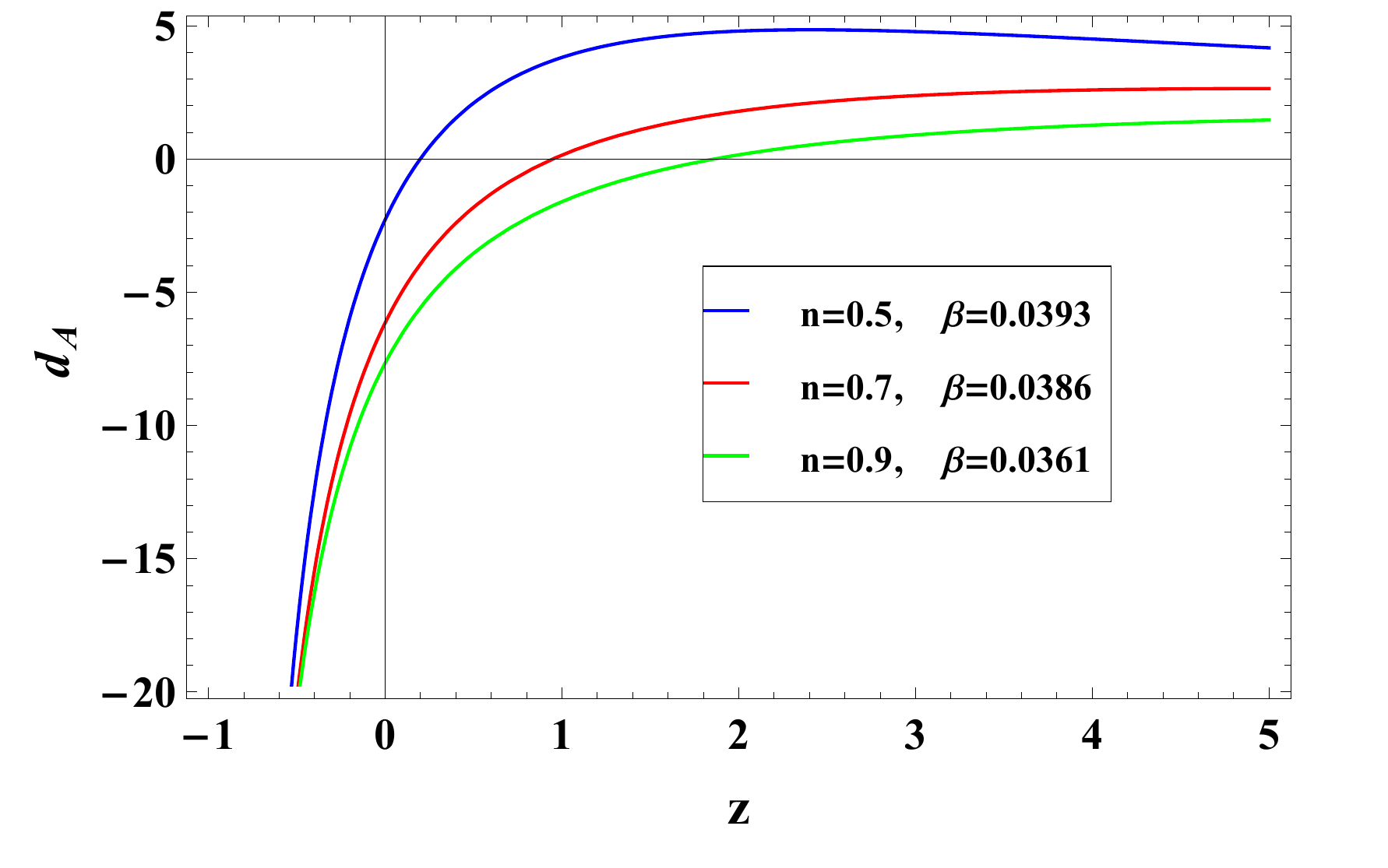}}
	\subfloat[]{\label{b}\includegraphics[scale=0.45]{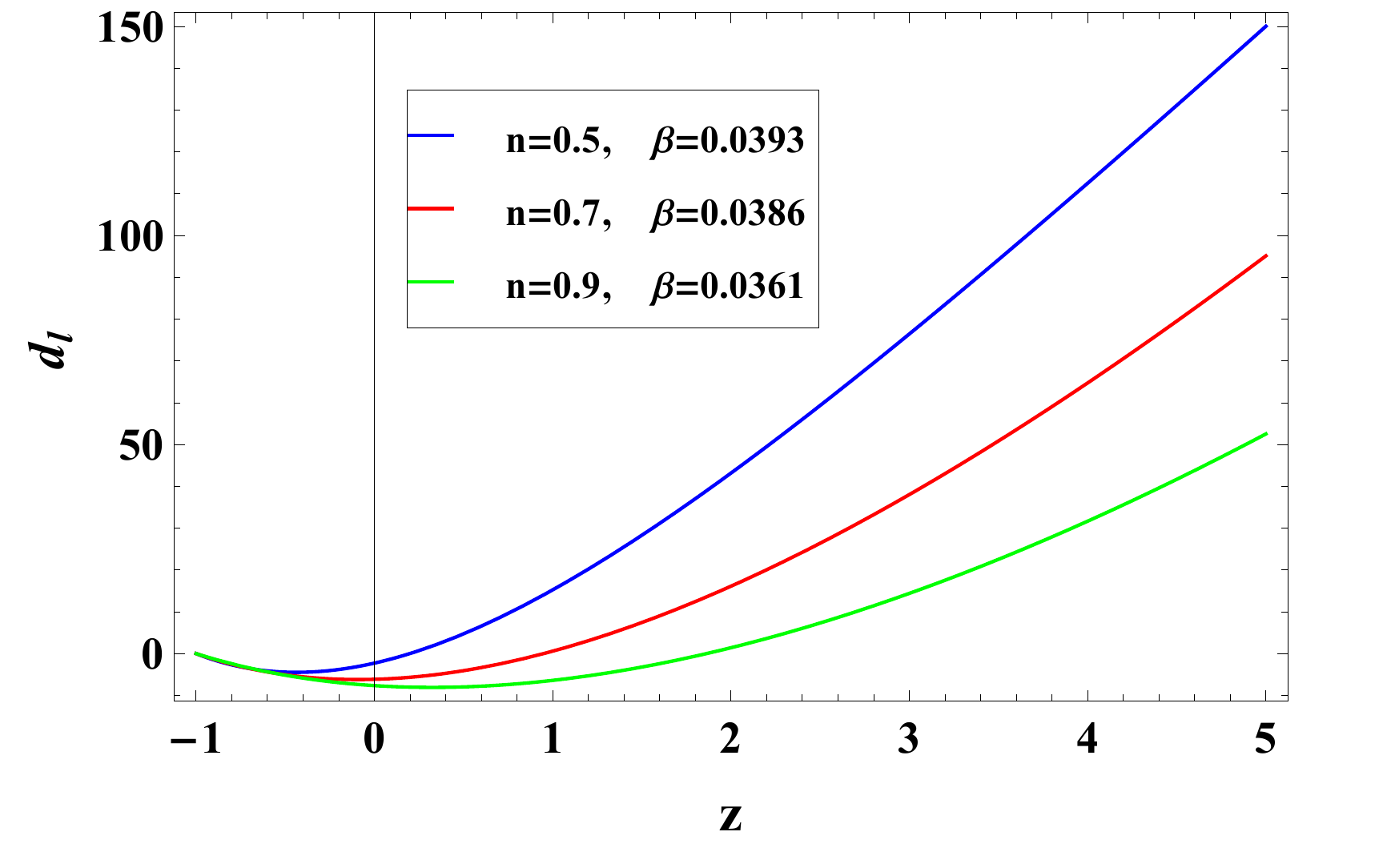}} 
	\caption{\scriptsize Depiction of graphs of $ d_{A} $ and $ d_l $  vs. $ z $.}
\end{figure}

\subsection{ Luminosity distance}

The gathering of distance versus redshift data for sources whose absolute luminosity is precisely known is a strong approach of exploring the cosmic expansion. The expansion background of the Universe can be determined by plotting this data \cite{Melia:2013sxa}. If we have a bunch of luminous sources with known luminosity $ l $ at different redshifts, we may compute the luminosity distance $ d_l(z) $ by monitoring the flux from these sources. The knowledge of $ d_l $,  can help in prediction of  the Universe's geometry. When compared to the cosmological constant dominated model, a given object at a specific redshift will look brighter in the matter dominated model. Though this appears to be a simple concept in theory, the statistical analysis proves to be fairly difficult \cite{Padmanabhan:2002ji}. The luminosity distance as a function of $ z $ is formulated as
\ba \label{33}
&& d_l=(1+z) ~d(z).
\ea
Using equation (\ref{30}) in (\ref{33}) we get
\ba \label{34}
&& d_l=(1+z) a_0 (t^{(1-n)} E_n(\beta t) - \beta ^{(-1+n)} \Gamma(1-n,13.8 \beta)).
\ea
The variations of luminosity distance $ d_l $ with respect to the redshift $ z $ is presented in Fig. 8(b) and it is observed that the values diminish as we move from the early Universe to the present time and finally end up in late times.

\section{ Conclusion}

\qquad We studied the late-time behaviour of Universe in the background of the flat FLRW metric in $ f(R,T) = f_1(R)+2f_2(T) $ gravity. The field equations are non-linear ordinary differential equations and are solved by parametrizing the scale factor $ a(t) $ in a hybrid form and then proceeding to analyze the obtained results, which describe the various stages of the evolution of the Universe. The model exhibits the point-type singularity. The volume of the model increases as $ t\rightarrow\infty $. The positive values of the Hubble parameter for all values of $ n $ considered for the present model indicate that the model is expanding. This model also exhibits phase transition from early deceleration to the late-times acceleration of the Universe. The energy density for this model is found to be monotonically decreasing with time. The energy density starts from infinity at the initial singularity and then reaches zero in late times. The pressure change with respect to time demonstrates that the present model depicts a structure formation phase with positive pressure at first and then acts as a dark energy model with negative pressure at later times, possibly causing the Universe's accelerating expansion. The model converges to $ \Lambda $CDM in late times.\\

The energy conditions parameters supporting that we have a quintessence dark energy model and parameters indicating the stability of this model are presented graphically. The energy conditions NEC and DEC hold good for $ n=0.5, 0.7, 0.9 $ but violates SEC in late times. The violation of SEC in late times agrees with the present observation of the accelerating Universe. The kinematic approaches to the model using jerk, snap and lerk parameters are discussed and are consistent with $ \Lambda $CDM in late times $ \forall n $. The present model shows quintessence type behaviour as the plots of $ Om(z) $ shows convexity \textit{w.r.t.} $ z $-axis. Also, it shows stable behaviour in early times but is highly unstable in late times. The $ s^*-r $ trajectory for different values of $ n $ traverses from the quintessence region and converges to $ \Lambda $CDM in late times. The $ q-r $ trajectory $ \forall n $ traverses from the decelerating region in early times to the accelerating region and converges to $ SS $, the Universe's Steady State Model in late times. The values of the lookback time, proper distance, angular diameter decrease as we move closer to the late Universe, whereas the values of luminosity distance decrease as we move from the early Universe to the present time and increase in the late Universe.\\ 

Thus after analysing the results, we observe that our model begins with a point-type singularity and behaves like a perfect fluid model in early times, an accelerated expanding quintessence dark energy model at present and converges to $ \Lambda $CDM in late times. 

\vskip0.2in 
\textbf{\noindent Acknowledgements} J.K.S. and  A.S. express their thanks to  Prof. Sushant G. Ghosh, CTP, Jamia Millia Islamia, New Delhi, India for fruitful discussions and suggestions. Authors also express their thanks to the referee for his valuable comments and suggestions.

\end{document}